\documentclass[journal]{IEEEtran}

\IEEEoverridecommandlockouts

\usepackage{amsfonts}
\usepackage{amssymb}
\usepackage[cmex10]{amsmath}
\usepackage{amssymb}
\usepackage{graphicx}
\usepackage{graphics}
\usepackage{cite}
\usepackage{subcaption}
\usepackage{dsfont}
\usepackage{url}
\usepackage{flushend}
\usepackage{multirow}

\usepackage{color}

\usepackage[normalem]{ ulem }
\usepackage{soul}

\newcommand{\F}{\mathbb{F}}

\newcommand{\A}{\mathcal{X}} 

\newcommand{\SNR}{{\bf  SNR}}

\newcommand{\prob}{\text{Pr}}
\newcommand{\defas}{\overset{\text{def}}{=}}

\begin{document}

\title{{Exploring and Experimenting with Shaping Designs for Next-Generation Optical Communications
\\  
}
}%

\author{Fanny Jardel, Tobias A. Eriksson, Cyril M\'easson, Amirhossein Ghazisaeidi, \\ Fred Buchali, Wilfried Idler,  and Joseph J. Boutros
\thanks{F.~Jardel (e-mail: fanny.jardel@nokia-bell-labs.com),  C.~M\'easson, and A.~Ghazisaeidi are with Nokia Bell Labs, Paris-Saclay, F-91620 Nozay, France. T.~Eriksson is with Quantum ICT Advanced Development Center, NICT, 4-2-1 Nukui-kita, Koganei, Tokyo 184-8795, Japan. F.~Buchali and W.~Idler are with Nokia Bell Labs, D-70435 Stuttgart, Germany.    J.J.~Boutros is with Texas A\&M University, 23874 Doha, Qatar.}
\thanks{Part of this paper had been presented at the {\em European Conference on Optical communication} (ECOC), Gothenburg, Sweden, 2017.}
}

\markboth{Submitted to Journal of Lightwave Technology}%
{Jardel \MakeLowercase{\textit{et al.}}: Exploring and Experimenting Shaping Tradeoffs ...}
%



\maketitle

\begin{abstract}
A class of circular 64-QAM that combines `geometric' and `probabilistic' shaping aspects is presented. It is compared to square 64-QAM in back-to-back, single-channel, and WDM transmission experiments. First, for the linear AWGN channel model, it permits to operate close to the Shannon limits for a wide range of signal-to-noise ratios. Second, WDM simulations over several hundreds of kilometers show that the obtained signal-to-noise ratios are equivalent to -- or slightly exceed -- those of probabilistic shaped 64-QAM.  Third, for real-life validation purpose, an experimental comparison with unshaped 64-QAM is performed where 28\% distance gains are recorded when using 19 channels at 54.2 GBd. This again is in line -- or slightly exceeds -- the gains generally obtained with probabilistic shaping. Depending upon implementation requirements (core forward-error correcting   scheme for example), the investigated modulation schemes may be key alternatives for next-generation optical systems. 
\end{abstract}


\IEEEpeerreviewmaketitle


\section{Introduction}

\subsection{Historical Notes} \label{sec:history}
In communication theory, {\em shaping} is the art of adapting a mismatched input signaling to a channel model by modifying the per-channel-use distribution of its modulation points. Efficient information transmission schemes  may use various shaping methods in order to increase spectral efficiency. Many of them have been investigated over the years,  from {\em nonlinear mapping} over asymmetric channel models or {\em many-to-one mapping}
\cite{Gallager1968} to {{optical}} experiments involving {{\em non-uniformly shaped}} QAM signaling.
 In particular, research efforts from the 70s towards the 90s derive conceptual methods to achieve {\em shaping gains} in communication systems. Following the advent of trellis coded modulation \cite{Ungerboeck1982}, a sequence of works  \cite{Calderbank1987,Forney1989a,
Calderbank1990, Fortier1992, Forney1992, Khandani1993, Laroia1994} present operational methods and achieve a large fraction of the 
ultimate shaping gain associated with square lattices. Trellis shaping or shell mapping are implemented in applications such as the ITU V.34 modem. Non-uniform input signaling for the Gaussian channel is further investigated in \cite{Kschischang1993, Betts1994}. While several shaping schemes are based on the structural properties of lattices \cite{Buda1989,Boutros1996, Loeliger1997, Forney2000, Erez2005}, 
 the interest in randomized schemes rose in the late 90s after the rediscovery of probabilistic decoding  \cite{Berrou1993,Gallager1963}. Multilevel schemes such as bit-interleaved coded modulation  \cite{Caire1998} offer flexible and low-complexity solutions  \cite{Imai1977, Zehavi1992}. 
 In the 2000s, several schemes have been investigated or proved to achieve the fundamental communication limits in different scenarios as discussed in \cite{McEliece2001,Soriaga2003,Gabrys2012,Ling2014,Mondelli2014}.  

In the last years, practice-oriented works related to optical transmissions have successfully implemented different  shaping methods, from many-to-one and geometrically-shaped formats to non-uniform signaling. The latter, more often called {\em probabilistic shaping} in the optical community \cite{Palgy2012,Bocherer2015,Schulte2016,Kramer2016,Boutros2017}, has perhaps received most attention. Various transmission demonstrations and record experiments using shaped modulation formats have indeed been reported as, e.g., in \cite{Buchali2015, Yankov2014, Beygi2014,Fehenberger2015,Ghazisaeidi2016,Chandrasekhar2016,Zhang2016,Theresa2017,Jardel2017}. 
 For illustration purpose, $65$Tb/s of operational achievable rate using state-of-the-art dual-band WDM technologies, partial nonlinear interference cancellation, and  non-uniform signaling are reported in \cite{Ghazisaeidi2016}.

\subsection{Implementations Constraints and Future Optical Systems} \label{sec:nextgen}

This work is motivated in part by the use of advanced QAM formats and in part by non-binary information processing. 
The investigated formats are neither restricted to non-binary architecture, nor specific to any information representation, nor even constrained by any coding/modulation method. Depending upon the application, different design criteria might be considered. {In particular, despite the induced complexity, several advanced channel models envisioned for next-generation optical systems require the use of {\em circular and possibly high-dimensional constellations}.}
In one example, nonlinear particularities of the optical {\em fiber channel} should be addressed. Due to the third-order nonlinear Kerr effect, the fiber channel  becomes {\em nonlinear} at optimum launch power for WDM transmission \cite{Agrawal2012}. The perturbation-based model  \cite{Mecozzi2000,Dar2013,Carena2014,Dar2016,Ghazisaeidi2017} 
 shows that specific characteristics such as the 4-th or 6-th order moments of the random input may be taken into consideration. 
In another important example, non-unitary and multi-dimensional channel characteristics may be addressed.  In particular, the work in \cite{Awwad2013,Dumenil2017,Dumenil2018} shows that rotation-invariant formats are instrumental whenever polarization-dependent loss happens. It indeed permits to attenuate or even eliminate the angle dependency when dimensional imbalance occurs, hence removing capacity loss due to angle fluctuation. In addition, spherical constellations may facilitate implementations of MMA-type (multi-modulus algorithm) of MIMO blind equalization. 
 %
Various other system criteria may also enter the picture. 
 A {\em matching} between channel physical model and transceiver architecture (in particular, receiver algorithms) is key to enable the ultimate transmission performance. A conventional receiver chain (comprising sampling, chromatic dispersion post-processing, MIMO equalization, phase and channel estimation, channel decoding and demodulation) that operates in a sequential manner is quite often sub-optimal. Joint processing may be required to preserve the sufficient statistics and 
improve the receiver performance. 
An implementation solution consists of using conventional non-binary information processing associated with 
matching signaling.

As various digital communication schemes {requiring non-square-QAM-based constellation} are candidates for next-generation optical applications, this paper aims at providing design guidelines for modulation formats.

\subsection{Outline of the Paper} 

This paper presents results originally reported  in \cite{Jardel2017}. It deals with an experimental study on the use of specific modulation formats with high spectral efficiency for long-haul communications. 
The WDM fiber channel has been historically approximated in the linear regime, or in the limit of short reach communications with short to mid-size constellations, by the standard additive white Gaussian noise (AWGN) channel model {encountered in communications theory} \cite{Shannon1948, Gallager1968}. This paper investigates efficient modulation formats defined on the complex plane that operate very close to the fundamental communication limits of the Gaussian model. They are further tested in more complete scenarios, including the simulation of long reach cases, and, finally, experiments that validate the modulation proposals. Note that, because this work deals with first guidelines for advanced signaling and multi-dimensional optical systems,  it does not, at first, consider system-dependent optical models such as the enhanced Gaussian noise (EGN) model \cite{Carena2014}.

\section{Shaping and Optical Communications}

\subsection{Setup and Notations}

\subsubsection{Channel Model}  
A crude approximation of the fiber channel under current coherent WDM technologies (involving PDM and mismatched architecture) is represented by the complex-valued AWGN channel model. This model is valid in ideal back-to-back scenarios and short-range transmissions. 
For characterizing future optical systems, the performance in the linear regime  remains central at the first order. In most real-life scenarios however, long range communications create different types of (intra, extra, noise) nonlinear interference for which perturbations on the solution of the Manakov equation \cite{Mecozzi2000,Dar2013,Dar2016,Ghazisaeidi2017}  
 may provide some insight into the design and analysis of efficient constellation. See also \cite{ Agrawal2012,Carena2014,Eriksson2016}. In this paper, shaping tradeoffs are first addressed in the idealized linear regime. They are later tested in the nonlinear regime by simulations and experiments. Formally,  the receiver is assumed to see, independently at each channel use, an overall additive white noise equivalent to a complex-valued random noise $Z=Z_1+\sqrt{-1}Z_2$ where the independent $Z_1,Z_2$ obey a real-valued zero-mean half-unit-variance Gaussian distribution. We model the random channel output by 
\begin{align*}
Y  =\sqrt{\text{\small \SNR}} X+Z,
\end{align*}
 whereby $X\in\A$ is the random input with probability $p_X$ and $\SNR$ the signal-to-noise ratio. In case of continuous and power constrained input alphabet, the capacity of the model is achieved by the Gaussian distribution and equals $\log(1+\text{\small \SNR})$.

\subsubsection{Coding and Modulation}
This paper investigates simple but efficient time-invariant modulation formats. A format is defined by the pair $(\A,p_X)$ composed of the input alphabet (constellation of points in the complex plane) $\mathcal{X}$ and the input distribution $p_X$. The input alphabet is a {\em codebook} with indexes formed by letters (denoted by $B$ or $S$) of the original information alphabet. Shaping in this paper is seen as the art of optimizing the transmission performance of a format with bounded entropy. Recall that, if the resulting constellations  asymptotically sample a Gaussian density that achieves the capacity $\log(1+\text{\small \SNR})$, then the spectral efficiency gets optimized. Non-uniform signaling is obtained in \cite{Kschischang1993} by letting $p_X$ follow the Maxwell-Boltzmann envelope (or any other distribution). It is called {\em probabilistic shaping} and sometimes {\em probabilistic constellation shaping} in the optical literature, which leads to distinguish between geometric and probabilistic shaping aspects of a format $(\A,p_X)$.  Optimal system performance is measured in terms of achievable rates. The mutual information between $X$ and $Y$ is denoted by $I(X;Y)$. This quantity operationally corresponds to coded-modulation: it is termed the {\em CM information rate}. For practical (often mismatched) systems,  we may operationally refer to the achievable rate associated with conventional estimation of the representation letter (bit or symbol). This quantity corresponding to bit (or symbol) MAP estimation is termed the {\em B-CM} (or {\em S-CM}, respectively) {\em information rate}. In many instances, it coincides with the classical bit-interleaved (or symbol-interleaved) coded-modulation BICM (or SICM, respectively) framework of \cite{Zehavi1992,Caire1998,Wachsmann1999,Guillen2008,Martinez2009} and is a particular case of generalized mutual information (GMI) \cite{Merhav1994,Ganti2000}.  Unless stated otherwise, the information source is represented by the random binary variable $B$. Random binary vectors can be equivalently represented as random symbols $S_i=S_i(B_1,\cdots,B_{m'})$. Random symbol vectors can be  equivalently represented as  random channel inputs $X=X(S_1,\cdots,S_{m})$. The  S-CM information rate is then given as $H(S_1,\cdots,S_m)-\sum_{i=1}^m H(S_i|Y)$ (and similarly for the B-CM rate). In practice, simple Riemann-based integration methods are used to compute the different information rates. More details on achievable rates are given in Appendix \ref{app:rate}.

\subsection{Square Quadrature Amplitude Modulation}

\subsubsection{Definition} Popular modulation formats are based on {\em Pulse Amplitude Modulation} (PAM) per quadrature, for which $P=2^m$ real-valued  points $x_1,\cdots,x_P$ are equally spaced and centered around $0$. The alphabet set is denoted by $P$-PAM and %
\begin{align*}
2^m\text{-PAM} & \propto \{-(2^{m}-1),\cdots,-3,-1,1,3, \cdots,2^{m}-1\}.
\end{align*}
In current practice, the $P$ points are associated with equal probabilities $p_X=\frac{1}{2^m}$. The choice $P=2^m$ enables simple bit labeling. For notational convenience, we use  
$\A = 2^m\text{-PAM}  \defas \{-\frac{2^{m}+1}{\sqrt{E(m)}}, \cdots, -\frac{1}{\sqrt{E(m)}},\frac{1}{\sqrt{E(m)}},\cdots, \frac{2^{m}-1}{\sqrt{E(m)}}\}$
 where $E(m)\defas \frac{2^{2m}-1}{3}$ is a constant that normalizes the total power to one.  The Cartesian product of two $P$-PAM alphabets is called (square) {\em Quadrature Amplitude Modulation} and denoted by $P^2$-QAM.

\setlength{\unitlength}{2pt}
\begin{figure}[ht]
\centering
\begin{picture}(135,111)
\put(0,-2)
{
\put(-48,-105){\includegraphics[width=443pt,height=640pt]{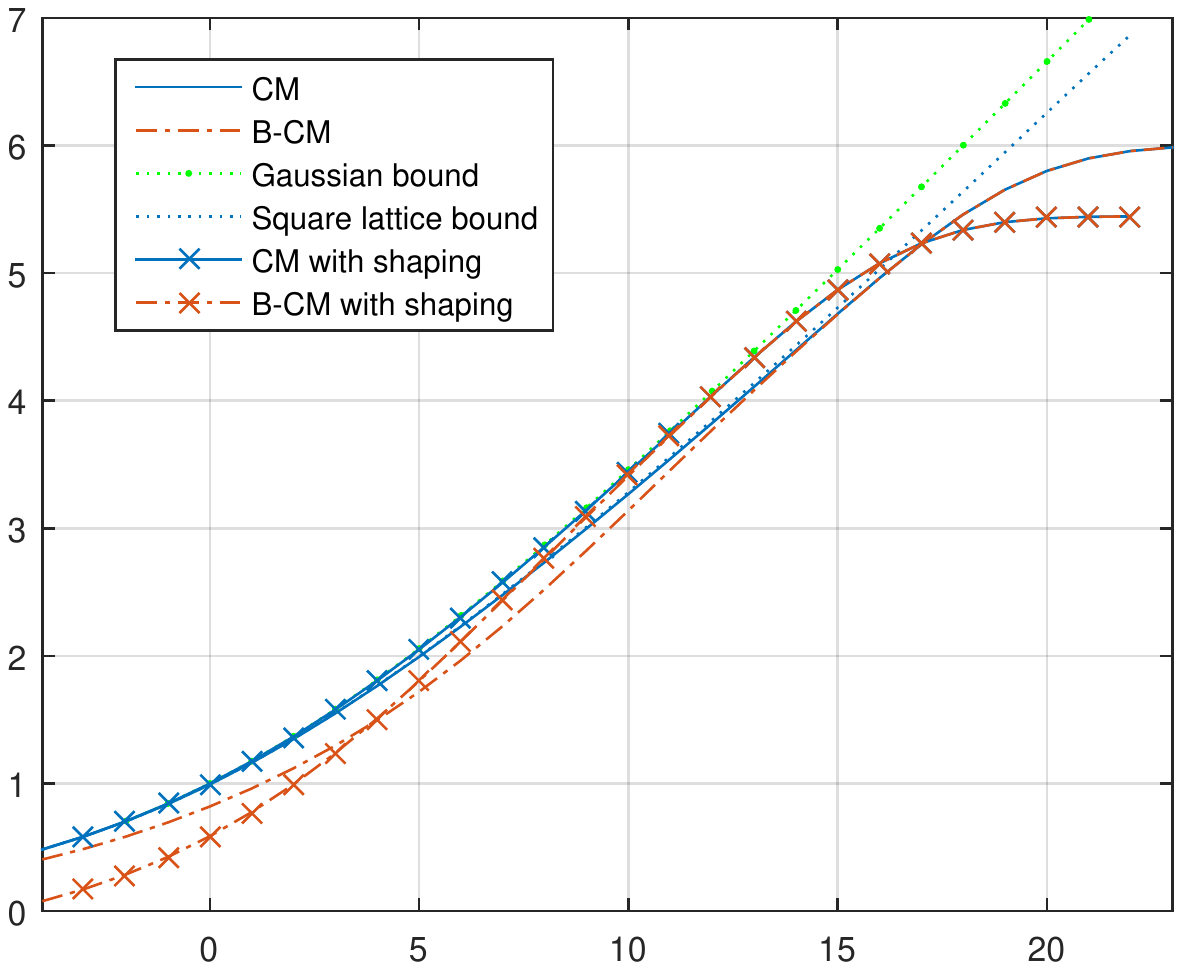}}
\put(-1,40){\rotatebox{90}{\footnotesize  Achievable Rate}}
\put(105,9){\rotatebox{0}{\footnotesize SNR (dB)}}
\put(103,102){\rotatebox{0}{\footnotesize$\longrightarrow~~~\longleftarrow$}}
\put(63,105){\rotatebox{0}{\footnotesize Ultimate SNR loss of $\frac{\pi \text{e}}6\approx1.53$dB}}
\put(67,45){\rotatebox{90}{\footnotesize$\longrightarrow~~\longleftarrow$}}
\put(74,55){\rotatebox{0}{\footnotesize Information loss due to}}
\put(74,51){\rotatebox{0}{\footnotesize square QAM input mismatch}}
\put(66,6){\rotatebox{90}{{\footnotesize  ~Shaping target SNR}~-~~~~~~~}}
}
\end{picture}
\caption{\footnotesize  
Achievable information rates for  square 64-QAM. The shaped distribution is optimized for a target SNR region around 10dB, which reduces the maximal transmitted entropy from 6bits per channel use down to 5.45. }\label{fig:qam64}
\end{figure}

\subsubsection{Properties} QAM formats are the constellations of choice in various communication systems. PAM  enables a natural Gray labeling of the information bits which increases performance  at mid-to-large signal-to-noise (SNR) ratios. Because I/Q QAM components remain independent in the presence of standard Gaussian noise, the statistical separation leads to  individualized demodulation schemes. Practical individual demodulation in this regime is enabled by the max-log approximation.  
Despite such important practical aspects, square QAM formats suffer from a noticeable drawback when associated with uniformly distributed codebooks. Geometric arguments on square lattices show that the overall transmission rate  is generally bounded away from the channel capacity \cite{Forney1989a}. In the case of additive Gaussian noise,  {\em shaping} permits to reduce this gap and asymptotically achieve up to $\frac{\pi \text{e}}6 \approx 1.53$dB of signal-to-noise ratio (SNR) gain. This is shown in Fig.~\ref{fig:qam64} for the example of 64-QAM. It can be observed that, when associated with an example of Gray mapping, the B-CM capacity deviates from the CM capacity at low SNR. As summarized in Section \ref{sec:history}, various shaping methods involving multi-dimensional geometric considerations have been devised in the past, e.g., shell mapping and trellis coded modulation for wire-line communications. In this paper, we focus on time-invariant non-uniform signaling \cite{Calderbank1990,Kschischang1993} as recently investigated in optical research (in particular in combination with {\em Probabilistic Amplitude Shaping} (PAS) \cite{Buchali2015,Bocherer2015,Ghazisaeidi2016}). This is exemplified for QAM in Fig.~\ref{fig:qam64} where non-uniform signaling is obtained using the Maxwell-Boltzmann distribution \cite{Kschischang1993}. 
In the target SNR region, the capacity of the shaped system approaches the ultimate limit given by the continuous Gaussian input distribution.
Beyond shaping loss, one may list additional drawbacks of square QAM that are specific to optical systems. Those include suboptimal equalization  in case of non-unitary impairments \cite{Awwad2013,Dumenil2017} or other mismatches as discussed in Section \ref{sec:nextgen}. Investigations on QAM-based variations are therefore  critical to envision alternative engineering designs.

\subsection{Circular Quadrature Amplitude Modulation}

\subsubsection{Definition}
Within this paper, we define a $q\times q'$-CQAM constellation to be a {\em circular} QAM format that is rotation-invariant in the I/Q plane. More precisely, by rotation of angle $\frac{2\pi}q$, the $P=q\times q'$ constellation points are  mapped onto constellation points with same associated probabilities. Examples include APSK formats as in \cite{Zhang2016}, or other constructions as in \cite{Larsson2017}. If $q=q'$, then a {\em $q^2$-circular quadrature amplitude modulation} ($q^2$-CQAM) is a two-dimensional constellation that includes $q$ {\em shells} (circles containing points of the same amplitude) with $q$ points per shell \cite{Boutros2017}. We write
\begin{align*}
q^2\text{-CQAM} & \propto\bigcup_{i=0}^{q-1}~e^{i\frac{2\pi}{q}\sqrt{-1}}~B, 
\end{align*}
where $B$ is a fundamental (connected or not) discrete set of $q$ points with distinct amplitudes. 
\subsubsection{Properties} One interesting aspect of CQAM-like formats is that they are naturally adapted 
to $q$-ary PAS coding. 
There are obviously many possible $q^2$-CQAM constructions. Depending upon the design criteria, e.g., the {\em figure of merit} \cite{Forney1989a} (minimum distance) as in \cite{Boutros2017}, different properties and  performance are obtained. In the sequel, we investigate different criteria options for CQAM constellation and perform specific optimization. We eventually focus on the CQAM construction of \cite{Boutros2017}. This particular CQAM construction, which originates from an exercise on the generalization of the PAS method, turns out to be particularly efficient with respect to CM capacity.  

\setlength{\unitlength}{1pt}
\begin{figure*}[h!]
\centering
\begin{picture}(900,360)
\put(0,65)
{
\put(9,0)
{

\put(-25,-80){\includegraphics[width=280pt]{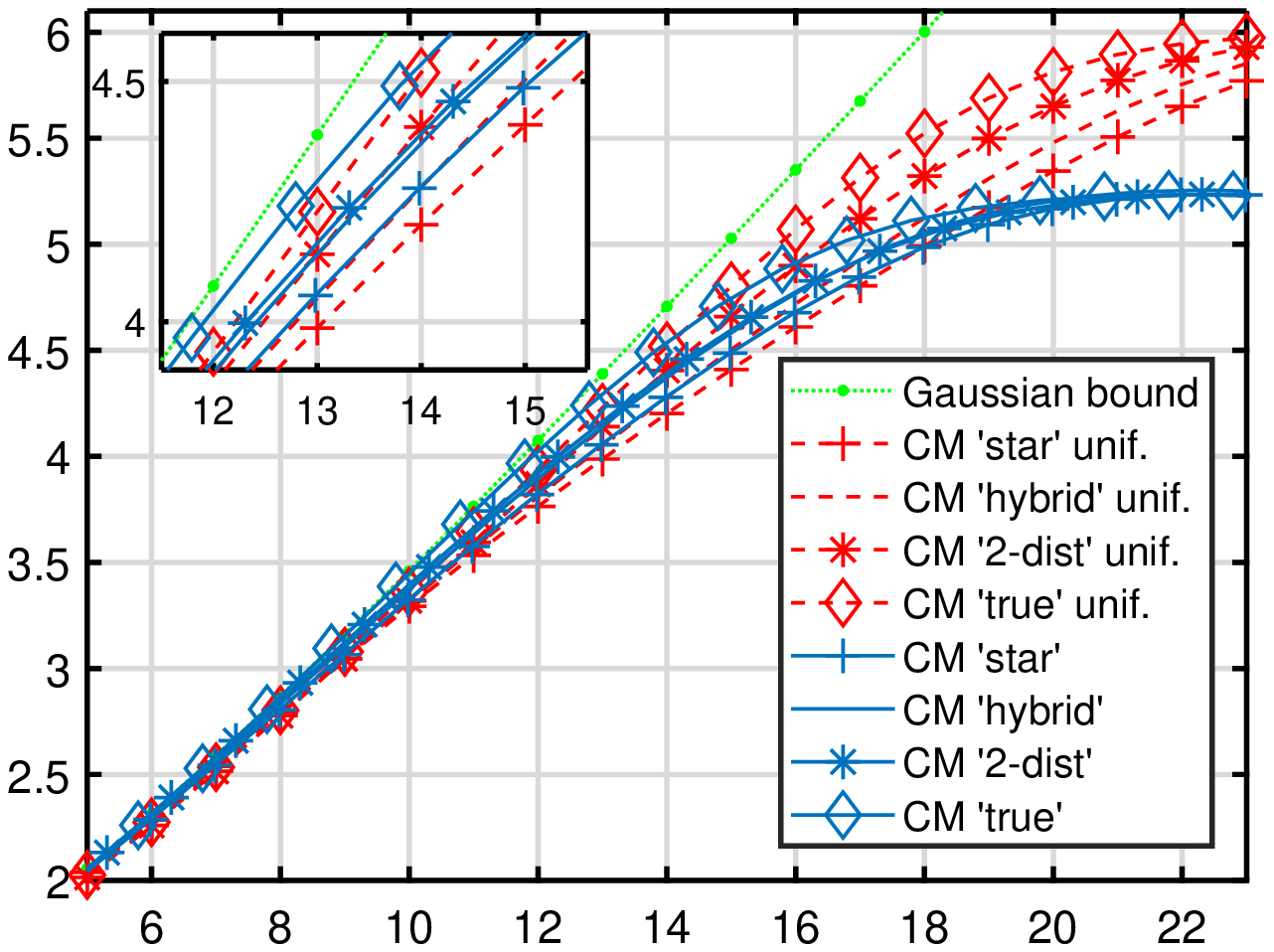}}
\put(-7,0){\rotatebox{90}{\footnotesize  Achievable Rate}}
\put(115,-70){\footnotesize(d)}

\put(185,-70){\rotatebox{0}{\footnotesize SNR (dB)}}
}
\put(-9,0)
{
\put(259,-80){\includegraphics[width=280pt]{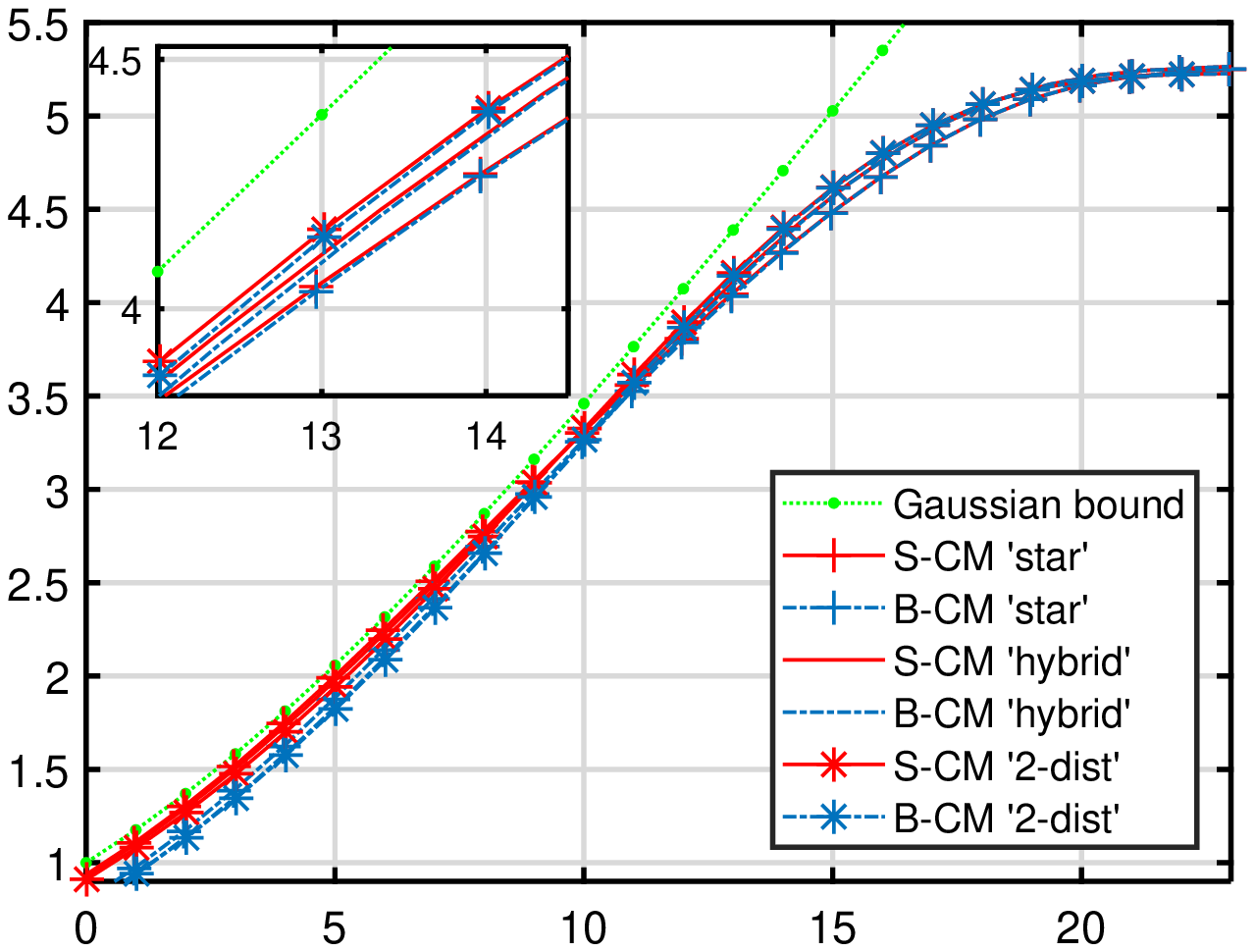}}
\put(282,0){
\put(-7,0){\rotatebox{90}{\footnotesize  Achievable Rate}}
\put(115,-70){\footnotesize(e)}
\put(195,-70){\rotatebox{0}{\footnotesize SNR (dB)}}
}
}
}
\put(0,54){
\put(-9,0)
{
\put(10,147){\includegraphics[width=172pt]{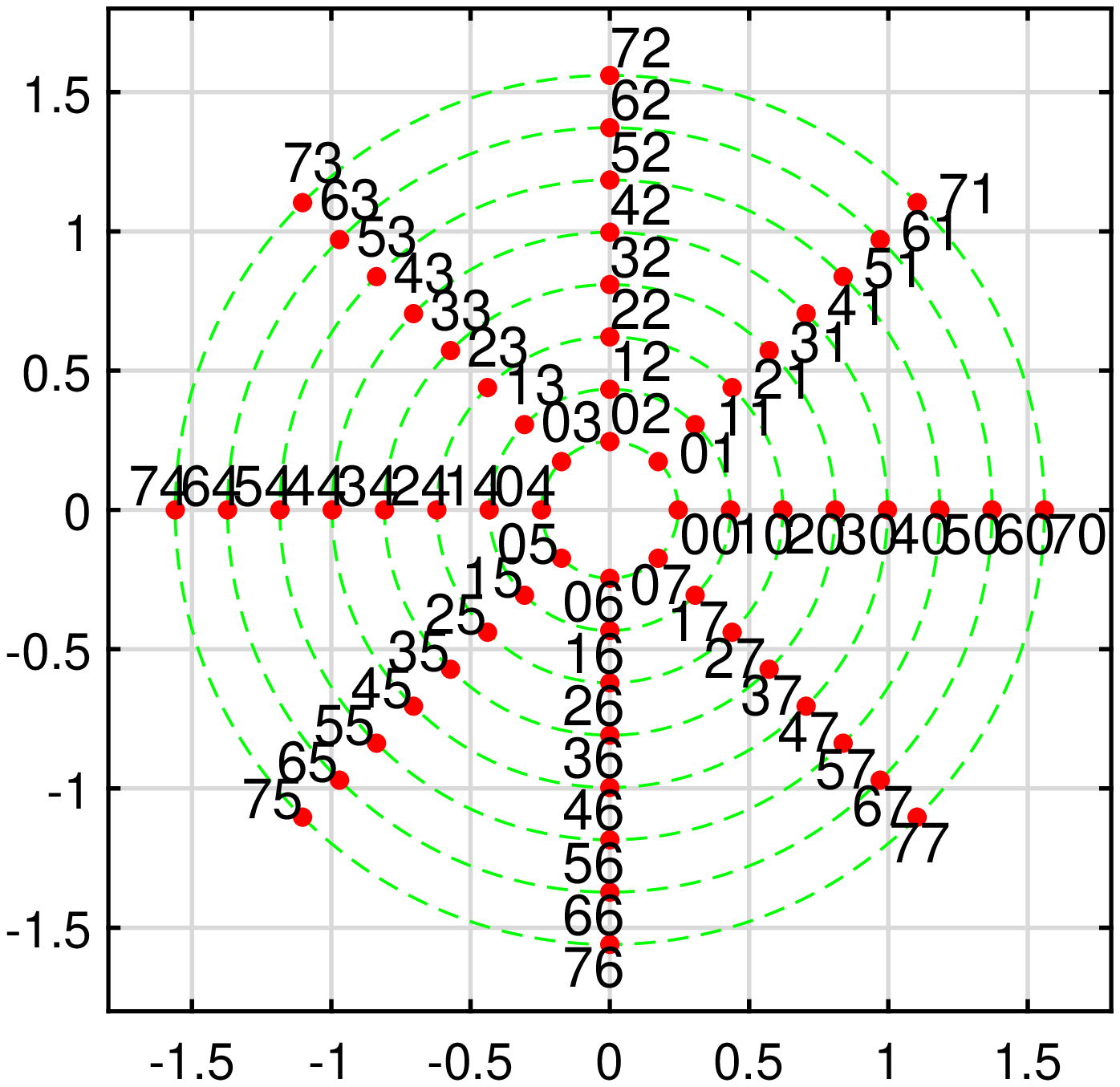}}
\put(160,162){\footnotesize `star'}
\put(89,140){\footnotesize(a)}
}
\put(158,0)
{
\put(16,147){\includegraphics[width=170pt]{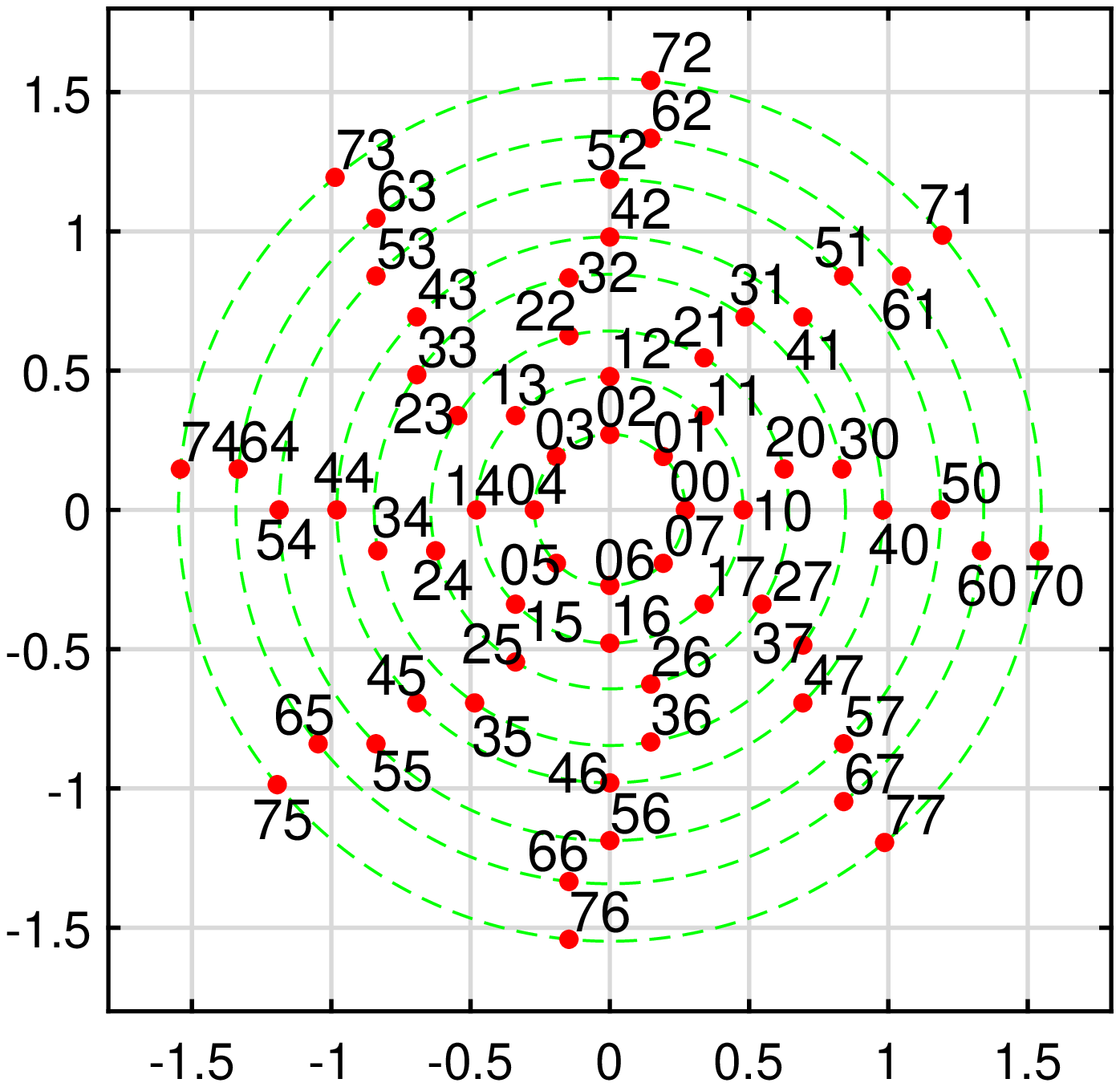}}
\put(155,162){\footnotesize `hybrid'}
\put(99,140){\footnotesize(b)}
}
\put(326,0)
{
\put(19,149){\includegraphics[width=174pt]{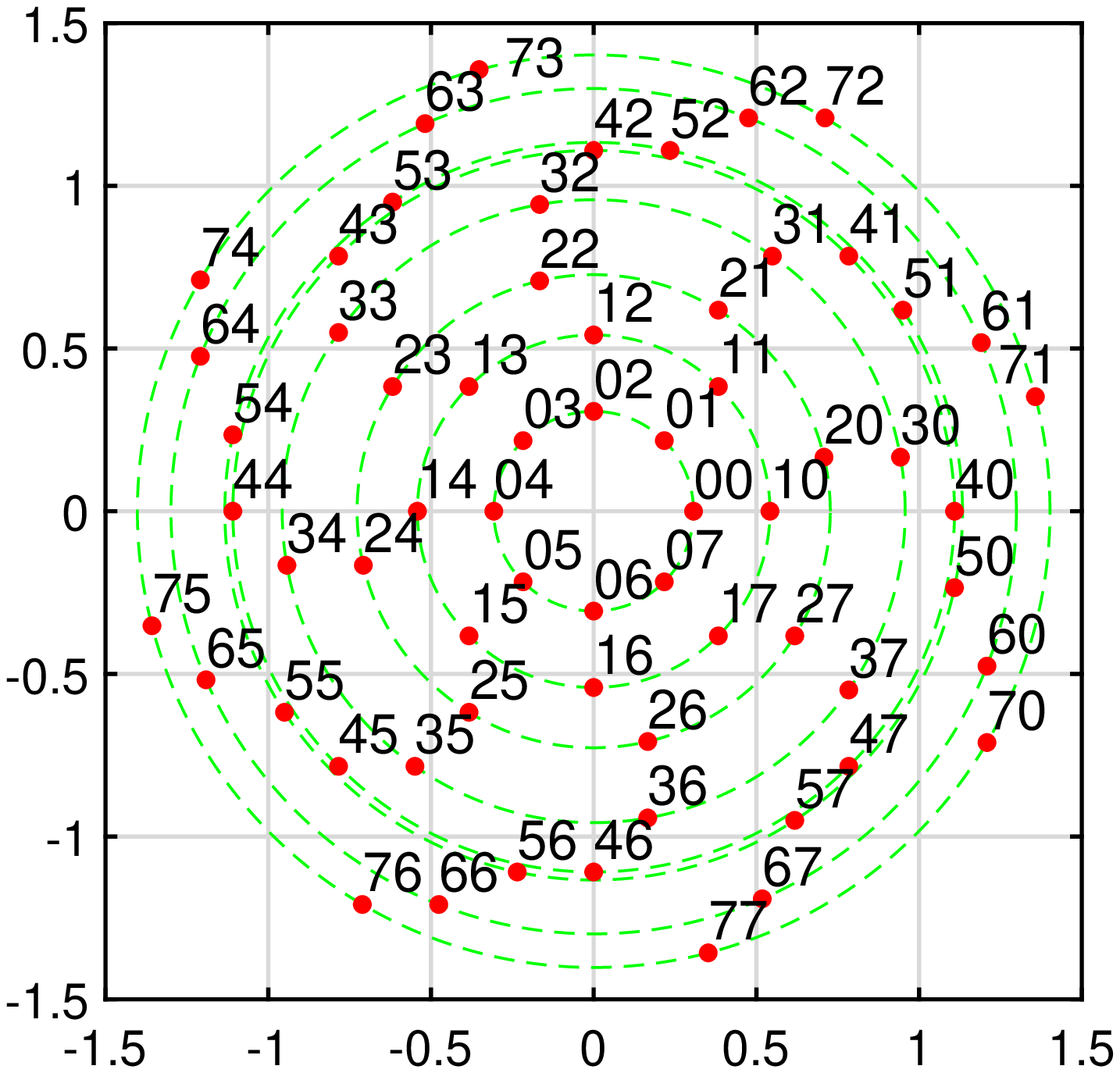}}
\put(161,162){\footnotesize `2-dist'}
\put(107,140){\footnotesize(c)}
}
}
\end{picture}
\caption{\footnotesize Types of CQAM-like constructions, information rates, and CM vs S-CM tradeoffs. The CM rate characterizes the optimal communication limits. The S-CM rate is a relevant operational quantity when working with $q=2^p$-ary-based architectures. The B-CM rate is given for the sake of completeness. Shaping parameters have been independently optimized for the 10dB SNR region while keeping an input entropy close to the shaped QAM of Fig.~\ref{fig:qam64}.
}\label{fig:cqamlike}
\end{figure*}

\subsection{Non-uniform QAM Signaling and the PAS method }

\subsubsection{Background} The reason PAS \cite{Bocherer2015,Schulte2016,Kramer2016} has been experimented with in optical communications are twofold. First, non-uniform signaling is obtained after shaping the source up front a ({possibly legacy}) coding system: this offers backward compatibility. Second, the distribution matcher (DM) provides an additional degree of freedom for rate adaptation: this may be a useful feature. The general PAS framework is found in Appendix \ref{app:pas}.
\subsubsection{Case with Square QAM }
 In its original binary instance, PAS is based on the antipodal symmetry of $2^m$-PAM. Indeed, up to power normalization, $2^m\text{-PAM}\propto \bigcup_{s=\pm 1}sB$, where $B=\{1, 3, \ldots, 2^{m}-1\}$.  Referring to Appendix \ref{app:pas}, the isomorphic representation 
\begin{align*}
2^m\text{-PAM}\equiv \{-1,+1\} \times B
\end{align*}
 permits to distinguish between signal amplitudes in $B$ and their sign in $\{-1,+1\}$. PAS in \cite{Bocherer2015} is based on the mapping $\{1,0\}\equiv\{-1,+1\}$ that encodes the sign while, independently, binary vectors label points in $B$. PAS is a layered coding scheme. The central channel coding layer uses a  linear code with systematic encoding and rate $R_{\text{C}}$ where  parity bits encode  amplitude signs. The systematic information is, for example, Maxwell-Boltzmann-shaped in a layer up front via, for example,  prefix-free source coding or similar methods: this is further used to encode the amplitudes at the end layer. For 64-QAM-based systems, the code rate constraint is $R_{\text{C}}\geq\frac23$.

\subsubsection{Case with Circular QAM}
A linear dense combination of $q$-ary symbols tends to asymptotically\footnote{The proof \cite{Gallager1963,Boutros2017} over $\F_q$ involves the $q$ roots of the unity as a generalization of the sign symmetry over $\F_2$. This generalization motivates the construction of CQAM over $\F_q$  with the use of {\em circular} symmetry \cite{Boutros2017}.} admit a uniform distribution \cite{Gallager1963,Boutros2017}. If PAS (with, e.g., standard LDPC, Turbo, or polar codes) tends to map uniformly-distributed parity bits into the signs of PAM points, then parity bits do not perturb amplitude shaping. The generalization of this property to alternative (non-binary) information representations is enabled by specific QAM format, among others $q \times q$-CQAM as in \cite{Boutros2017} when the underlying alphabet is assumed to be a finite field $\F_q$ with a prime $q>2$. A generalized PAS framework is presented in Appendix \ref{app:pas} where it is observed that the new schemes relax the code rate constraint to $R_\text{C}\geq \frac12$ for any $q$.  Referring to Appendix \ref{app:pas}, we use the isomorphic representation 
\begin{align*}
q^2 \text{-CQAM}\equiv \F_q \times B,
\end{align*}
where $B$ represents the fundamental region. In \cite{Boutros2017}, the main goal is to explore the use of $q$-ary codes by generalizing PAS and the binary sign flipping technique to the $q$-ary case. In this  paper, the goal is slightly different. For practical reasons, we are restricted to computation fields of characteristic $2$ and, in particular, $q=2^3$. As this field is an extension of the binary field, the nature of code constraint is less stringent and, even for PAS, suboptimal schemes with binary codes could be envisioned. The code rate constraint of the generalized framework however remains valid and of interest. For 64-CQAM-based systems, it is $R_{\text{C}}\geq \frac12$.

\begin{figure*}[ht]
\centering
\begin{picture}(500,165)
\put(0,0)
{
\put(-25,0){\includegraphics[width=220pt]{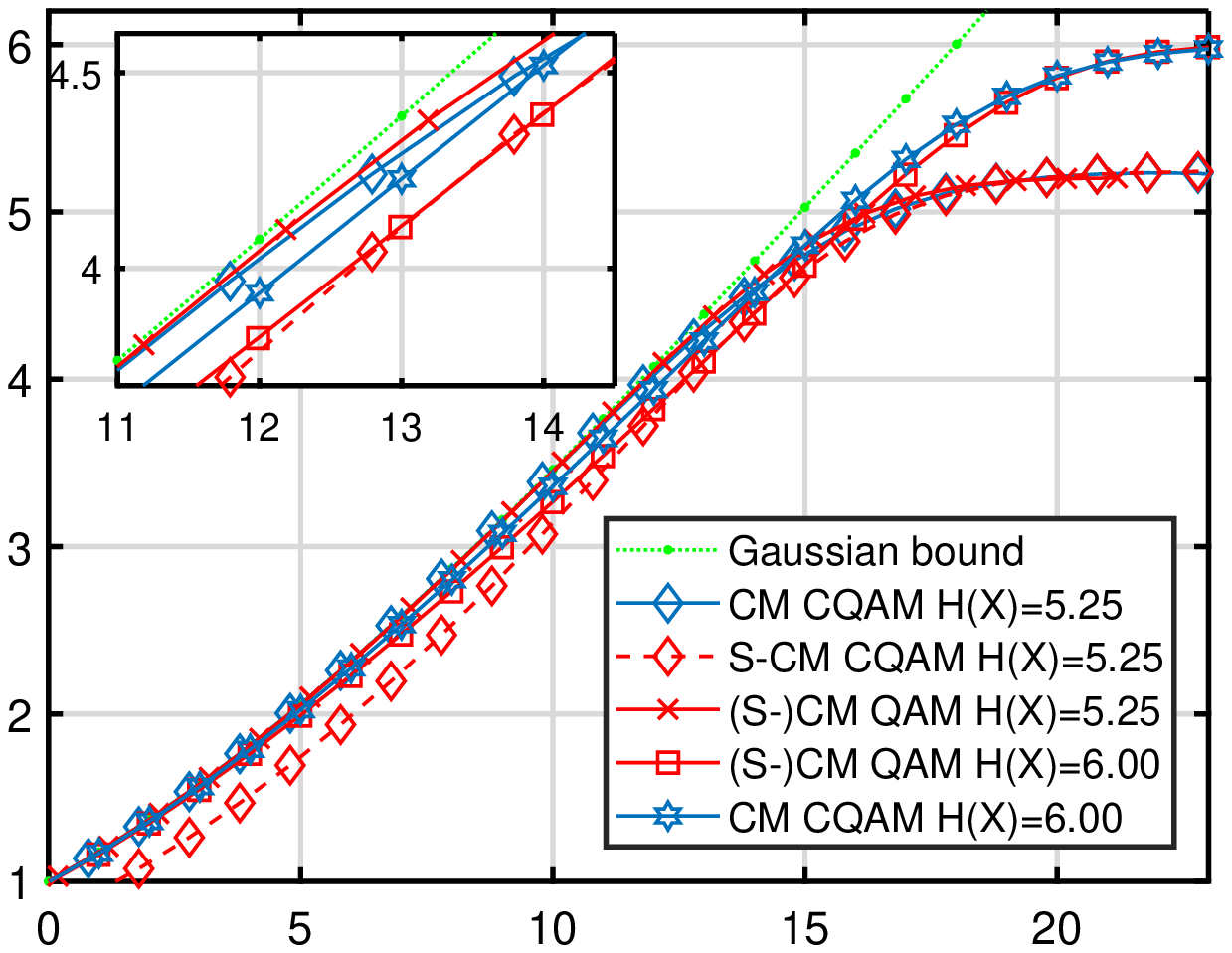}}
  
\put(-9,60){\rotatebox{90}{\footnotesize  Achievable Rate}}
\put(64,7){\rotatebox{0}{\footnotesize SNR [dB]}}
\put(78,-2){\rotatebox{0}{\footnotesize (a)}}
} 
\put(0,0)
{
\put(188,14){\includegraphics[width=149pt]{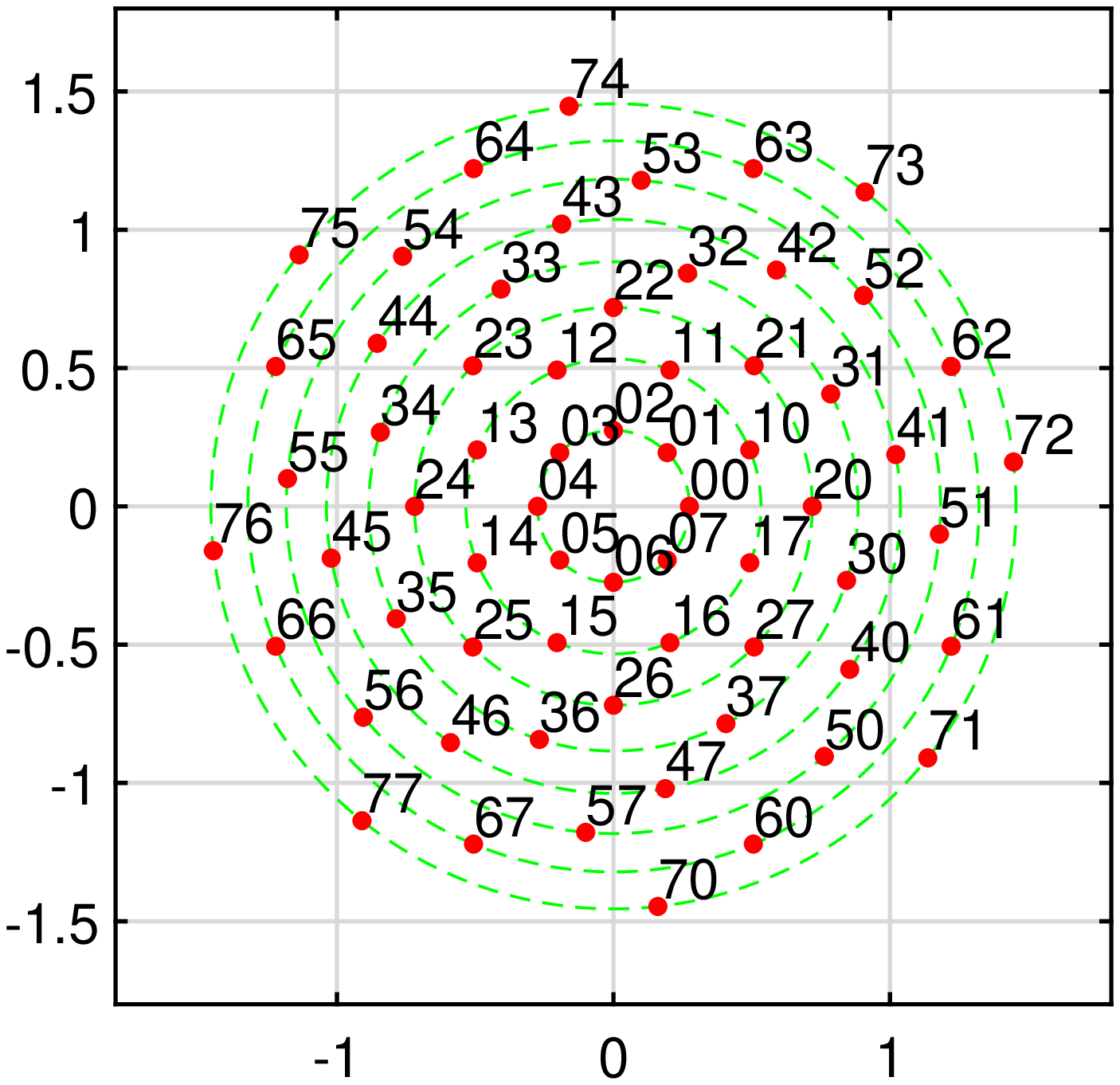}}
\put(260,-2){\rotatebox{0}{\footnotesize (b)}}
\put(190,5){\rotatebox{0}{\footnotesize Experimental `true' $8\times8$ CQAM \cite{Boutros2017,Jardel2017}}}
}
\put(0,0)
{
\put(350,10){\includegraphics[width=154pt]{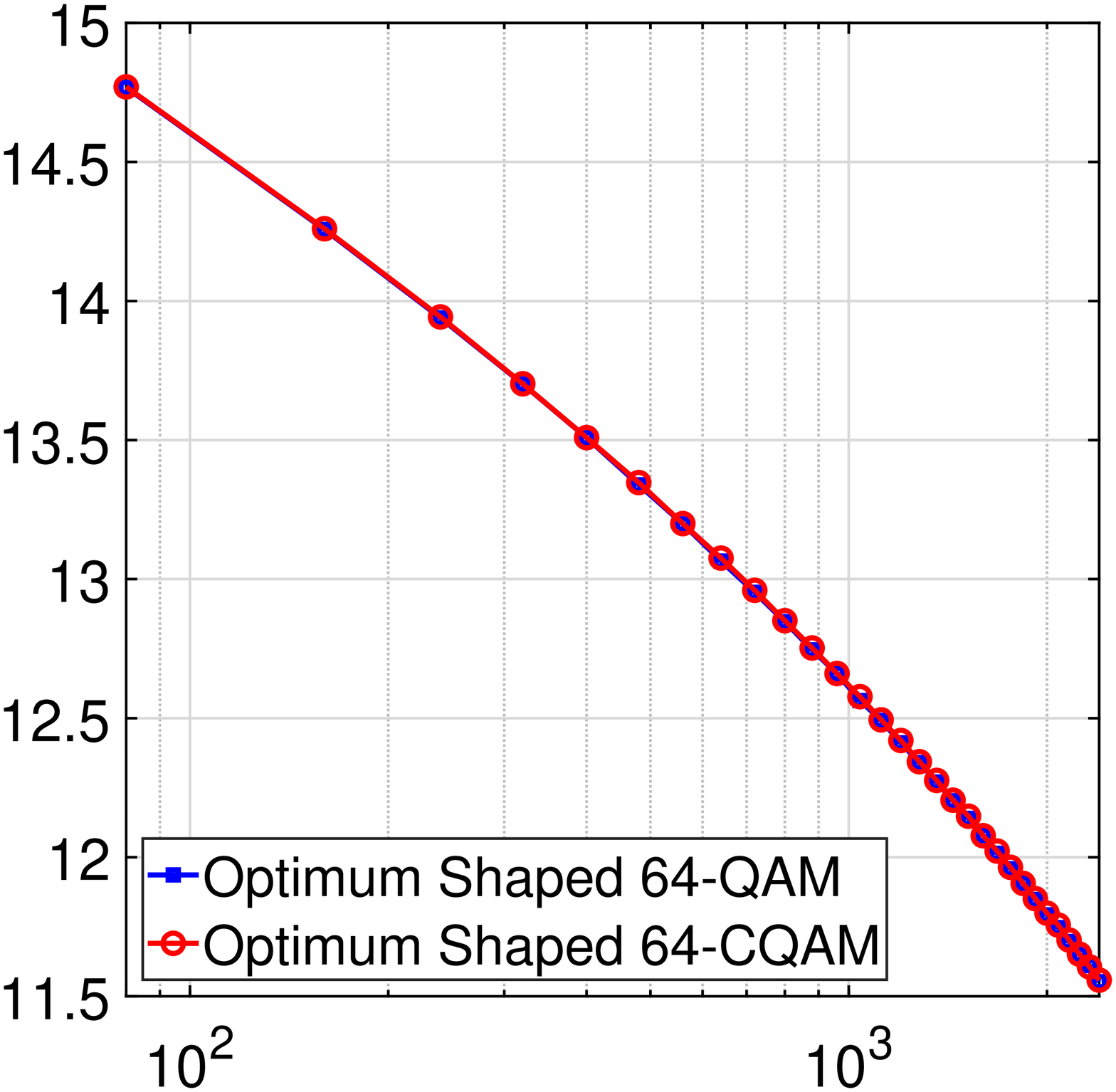}}
\put(342,75){\rotatebox{90}{\footnotesize  SNR [dB]}}
\put(410,7){\rotatebox{0}{\footnotesize Distance [km]}}
\put(420,-2){\rotatebox{0}{\footnotesize (c)}}
}
\end{picture}
\caption{\footnotesize  (a) $\&$ (b) Experimental $2^3\times2^3$-CQAM (CM-optimized in the SNR region below 12dB). Information rate as a function of SNR over the linear AWGN model. 
(c) Obtained SNR as a function of transmission reach for shaped 64-QAM and shaped 64-CQAM. }\label{fig:cqam}
\end{figure*}

\subsubsection{Perspective} 
Notice that the non-uniform signaling formats presented in  this paper may serve as general baseline performance guidelines. They are not PAS-specific guidelines. The obtained results and insights  can be used in a variety of 
 coding system models and coded modulation schemes.
As discussed in the previous section, this paper is an experimental validation of efficient modulation formats taking first into account advanced core linear model constraints, in particular multi-dimensional non-unitary polarization-multiplexing and multi-ary DSP implementations. Non-linear interference or other mismatches that depend on the transmission distance are treated in a second phase. Because optimizing the Euclidean distance is sufficient in the simplified high SNR linear scenarios and because we target low-complexity practical solutions, we use Maxwell-Boltzmann distributed amplitudes as the implied distorsion is negligible. Notice that alternative rate-distortion methods (e.g., Blahut-Arimoto) to optimize the input pair $(X,p_X)$ have been recently investigated. In \cite{Yankovn2017} and \cite{Renner2017}, the mutual information is optimized within the framework of the EGN model \cite{Carena2014,Fehenberger2016, Pan2016} (the classical one-dimensional Gaussian model with non-linear noise discussed in the optical literature) or the split-step Fourier transform.

\section{Modulation Tradeoffs}

\subsection{Geometric Shaping via Circular QAM Formats}

The general construction of $q^2$-CQAM constellations consists in populating $q$ shells with $q$ points that are uniformly distributed on the $q$-th circle. A construction criterion permits to control spacing and phase-offset between shells. The selection of a particular criterion may be motivated by {\em geometric} considerations. It may also be combined with the further optimization of the transceiver design at a given target SNR under {\em non-uniform} signaling. Hence, for standard receiver architectures, the construction and the choice of Maxwell-Boltzmann parameters may aim at maximizing the B-CM capacity under suboptimal bit-based estimation. For evolved architectures or advanced fiber channel models, they may be chosen to maximize the CM capacity. Recall that the latter case is considered in \cite{Boutros2017} and experimented with in \cite{Jardel2017}; it is indicated as the `true' CQAM reference in Fig.~\ref{fig:cqamlike} where several CQAM-like examples are depicted. Let us describe them. 

{\em Example 1: `Star' Construction.} 
Fig.~\ref{fig:cqamlike}a depicts `star-like' CQAM. Such constellations are similar to APSK constellations used in \cite{Zhang2016}. Their geometry combined with Gray mapping make them perform well in practice. In Fig.~\ref{fig:cqamlike} we use a shell spacing equal to $d_{\text{min}}$. A Gray mapping is represented using pairs ($m=2$) of labels in $\F_{2^3}=\{0,1,\cdots,7\}$. Respective labels of $\{0,1,\cdots,7\}$ are represented using the binary alphabet 
$\{111,110,100,101,001,000,010,011\}$. This translates into a (non-optimized) binary Gray code for the CQAM constellation where each point is now labeled by $m=6$ bits.

{\em Example 2: `2-dist' Construction.}   
Fig.~\ref{fig:cqamlike}c represents a CQAM constellation that has been constructed based on a `two-distance' criteria. The greedy construction is performed with respect to $d_{\text{min}}$ and the second minimum Euclidean distance given the first two shells (the second at $d_{\text{min}}$ being a scaled version of the first). Compared with a `star' design, this naturally increases  the CM capacity while attempting to maintain good properties for Gray labels.

{\em Example 3: `Hybrid' Construction.} 
Fig.~\ref{fig:cqamlike}b  represents a balance between the previous two example. 
 Angular regions have been preserved in addition to the 
 two-distance criteria. Depending upon receiver design and available information, angular region and mapping can be adapted for optimizing bit-based estimation performance, see also \cite{Essiambre2010}.

From these examples, we see that, if conventional (mismatched) BICM estimation were to be assumed, then a tradeoff may have to be made as the respective behaviors of CM and S-CM capacities are reversed in the operational SNR region. Indeed, while similar shaping and maximal input entropy have been represented, it appears that the performance behavior is first conditioned by the initial geometric properties of the constellation, then enhanced by a particular set of shaping parameters. For the CM capacity, it is known that minimum Euclidean distance maximization leads to remarkable CM capacity at high SNR. When based on that criterion, CQAM appears to be very efficient. Notice however that, if conventional (mismatched) BICM estimation is to be used for technical reasons, this cannot be fully exploited and a tradeoff may have to be made.  
In the remainder of this paper, we focus on CQAM constructions that, when shaped, maximizes the CM capacity. 

\begin{figure*}[!t]
\begin{center}

\includegraphics[width=1.9\columnwidth]{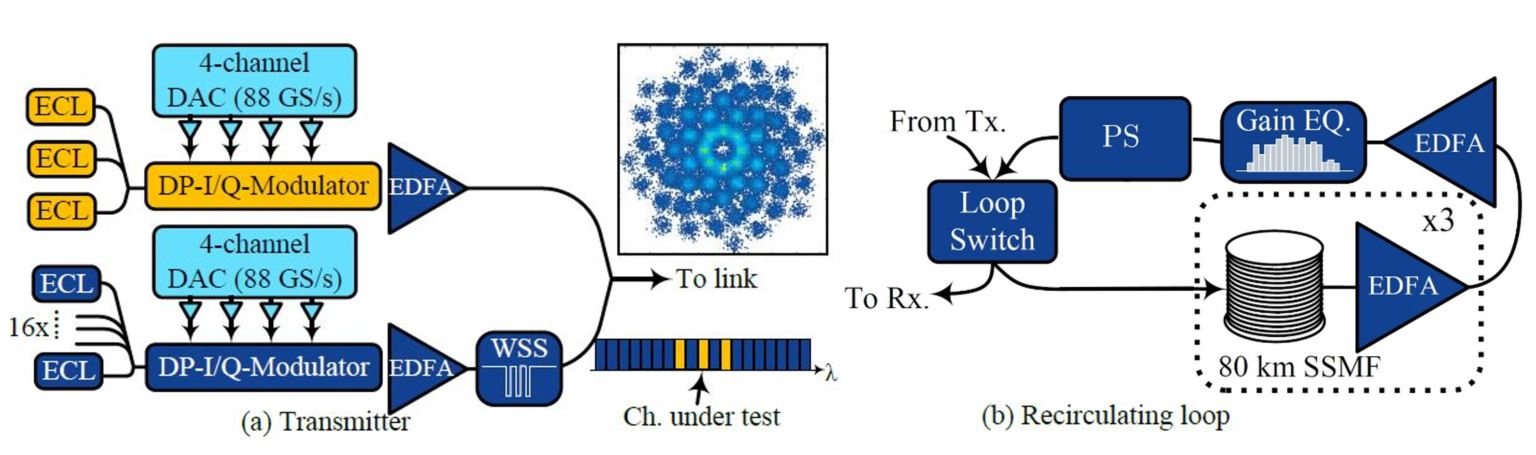}
   \caption{\footnotesize Experimental Setup.}
   \label{fig:tx}
   \end{center}
\end{figure*}

\subsection{Optimization for CM Capacity and Linear AWGN Model}

Let us present in more details the type of CQAM-like constructions introduced in \cite{Boutros2017}. It is solely based on the minimum Euclidean distance and is referred to as $q^2$-CQAM in the sequel. The chosen criteria maximizes the {\em figure of merit} or the ratio between ${|\A|d_\text{min}^2(\A)}$ and ${\sum_{x\in\A}|x|^2}$, where $d_\text{min}^2(\A)=\min_{(x,x') \in \A^2, x\ne x'} |x-x'|^2$ denotes the minimum squared Euclidean distance.  In practice, the minimum distance of the power-normalized constellation $\A$ is first maximized via a greedy procedure. Then, Maxwell-Boltzmann shaping is performed such that, for a given SNR, the gap between the CM information rate denoted by $I(X;Y)$ and the ultimate limit denoted by $\log(1+\SNR)$ is minimized. An optional stretching step may be performed, see \cite{Boutros2017}. This optimization procedure is sufficient to devise optimized constellations very close to the Shannon bounds of the core model. More importantly, this simple optimization is guided by operational constraints, i.e., the construction of circular constellations. 
 The $q^2$-CQAM format that supports the experimental work conveyed in this paper is represented in Fig.~\ref{fig:cqam}b. In terms of CM capacity, the stretched version achieves performance that are less than 0.1dB away from the Gaussian capacity $\log(1+\SNR)$. This is illustrated by Fig.~\ref{fig:cqam}a where the optimization has  been done for target SNRs around 10dB as slightly above. It can be seen that shaped 64-CQAM and shaped 64-QAM have similar performance at the operating point. 
Notice that these observations concern the CM capacity. The receiver architecture may require specific demapping nodes and estimation methods to take full benefit of the circularly-symmetric format. 

\subsection{Optimization for Nonlinear Long-Haul Communications}

The values of spectral efficiency of shaped constellations (whether square QAM or circular QAM) are very close to the Shannon capacity for the AWGN channel. For long-distance transmissions however, their performance are tested and reevaluated in the presence of the fiber nonlinear impairments. As previously mentioned, this is justified as the 4-th and 6-th moments of the constellations appear in the expressions of the total effective SNR. By {\em total effective SNR,} we mean the SNR of the sampled received signal assuming symbol-by-symbol coherent detector, without nonlinear equalization. In this case, the nonlinear distortions are effectively considered as additive Gaussian noise, but with a variance that scales cubically with channel average launched power, and depends on constellation moments \cite{Ghazisaeidi2017}.
We compared the performance of 64-CQAM and shaped 64-QAM formats for a nonlinear fiber channel using the theory presented in \cite{Ghazisaeidi2017} (see Eq.~(123) therein). We assumed a system with 19 channels, modulated at 54.2 GBd, and spaced at 62.5 GHz. The link consisted of 80 km spans of SMF fiber. We theoretically computed the SNR of the central (i.e., the 10-th) channel at the receiver side at the nonlinear threshold (i.e., the optimum launched power that maximizes the SNR) as a function of the number of spans. This is given for each of the two modulation schemes, and for a given span count. Referring to \cite{Kschischang1993, Bocherer2015}, the parameter of the Maxwell-Boltzmann of the PMFs of each scheme varied between 0 and 4, and the probability distribution that maximized the optimum SNR is found for each constellation by exhaustive search. 
Fig.~\ref{fig:cqam}c illustrates the optimized SNR vs distance for both formats. We observe that the two modulation schemes have very similar performance in the nonlinear regime. 
This observation permits us to assert that the shaping scheme proposed in this work is (at least equally) as robust as the existing solutions to fiber nonlinear impairments. 

\section{Experimental Setup}

\begin{figure}[!h]
\begin{center}
\includegraphics[width=1\columnwidth]{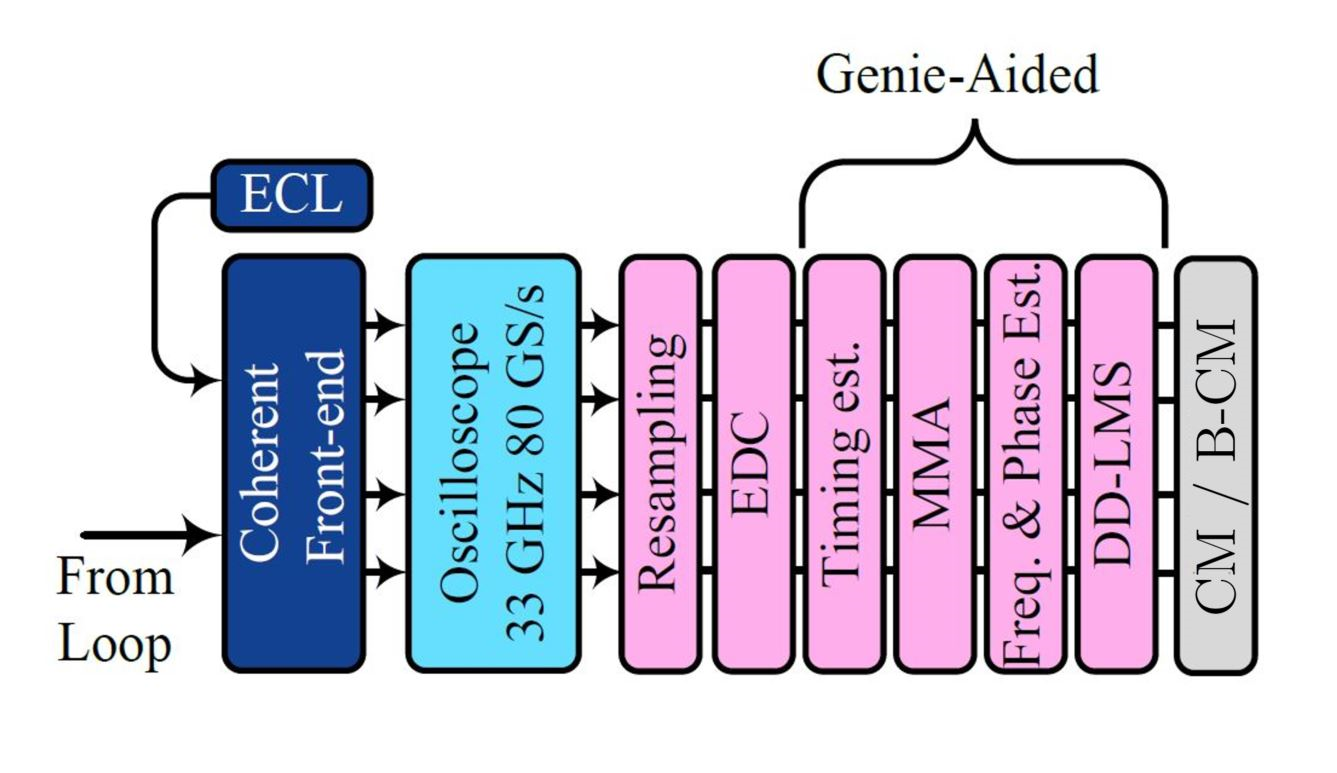}
   \caption{\footnotesize Receiver.}
   \label{fig:tx2}
   \end{center}
\end{figure}

The experimental setup is shown in Fig.~\ref{fig:tx}a. The transmitter is based on two four-channel digital-to-analog converters (DACs) running at 88~GS/s generating 54.2~Gbaud polarization multiplexed 64-QAM or 64-CQAM, using raised cosine pulses with a roll-off factor of 0.08. The length of the random transmitted sequences are 184320 symbols. In total, we modulate 19 WDM channels with a channel separation of 62.5~GHz using external cavity lasers (ECLs) with linewidths of around 100~kHz. One DAC is used to generate the channel under test and its two second nearest neighboring channels. The second DAC generates the remaining 16 channels. Independent symbol patterns are used for the two DACs. After the dual-polarization I/Q-modulators, we use  erbium doped fiber amplifiers (EDFAs) to  boost  the  signal.
 In the loading channel arm, we use a wavelength selective switch  to  remove the  in-band  amplified  spontaneous emission noise for the channel under test before combining the signals from the two transmitters.
The signals are either noise loaded and detected in a back-to-back scenario, or transmitted over the recirculating loop depicted in Fig.~\ref{fig:tx}b. The recirculating loop consists of three spans of conventional single mode fiber (SSMF), EDFAs and a polarization scrambler (PS). A programmable gain equalizer is used to equalize the power of the WDM channels and to filter out the ASE noise that is outside of the total channel count. The  signals are  detected using  a conventional  polarization diverse coherent  receiver, shown in Fig.~\ref{fig:tx2} and  digitized  using a  33~GHz  80~GS/s  real-time sampling oscilloscope. 

For a fair comparison between 64-CQAM and 64-QAM without penalty due to potential suboptimal equalization, we use a genie-aided-based digital signal processing (DSP) solution. Notice that, in practice, for future system implementations, pilot-aided DSP solutions are proved to be efficient \cite{Ivan2017}. Phase recovery follows from simple inverse mapping and standard DSP techniques are applied. 
The DSP starts with resampling to 2 samples/symbol followed by electronic dispersion compensation (EDC). Timing estimation, as well as polarization demultiplexing and adaptive equalization using a multi-modulus algorithm (MMA) is applied where knowledge of the transmitted data is used to calculate the error function. The signals are then sent to a frequency offset estimation and phase estimation stage. Finally, in this experimental demonstration, a symbol-spaced real-valued decision-directed least mean square (DD-LMS) equalizer is used independently on the signals in the x- and y-polarization to compensate for any remaining imperfections such as transmitter side timing skew. The parameters of the genie-aided DSP 
are adapted 
such that the performance is close to that of blind DSP for 64-QAM. To assure a fair comparison, the same parameters are then used for 64-CQAM. 

\begin{figure*}[!t]
    \begin{minipage}[h]{0.5\linewidth}  
    \hspace{-0.15cm}
         \put(20,5){\includegraphics[width=0.9\columnwidth]{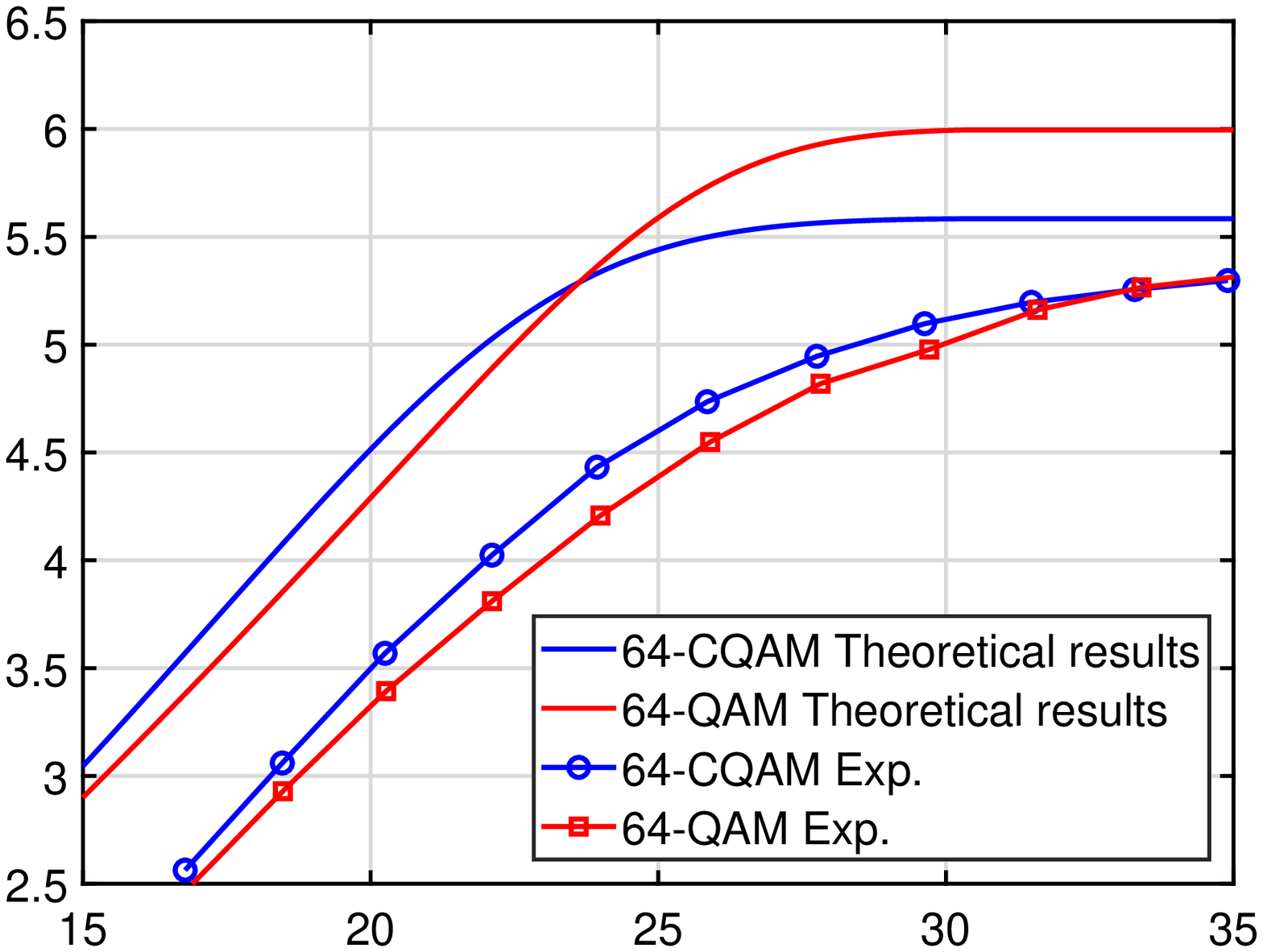}
         \put(-240,60){\rotatebox{90}{\footnotesize  MI [bit/2D-symbol]}}
         \put(-140,-10){\rotatebox{0}{\footnotesize OSNR [dB] 0.1 nm}}}
         
    \caption{\footnotesize Exp. back-to-back results. \hspace{-1cm} \label{b2b}  }
        \end{minipage} \hfill
         \begin{minipage}[h]{0.5\linewidth}  
         \hspace{-0.15cm}
           \put(20,5){\includegraphics[width=0.9\linewidth]{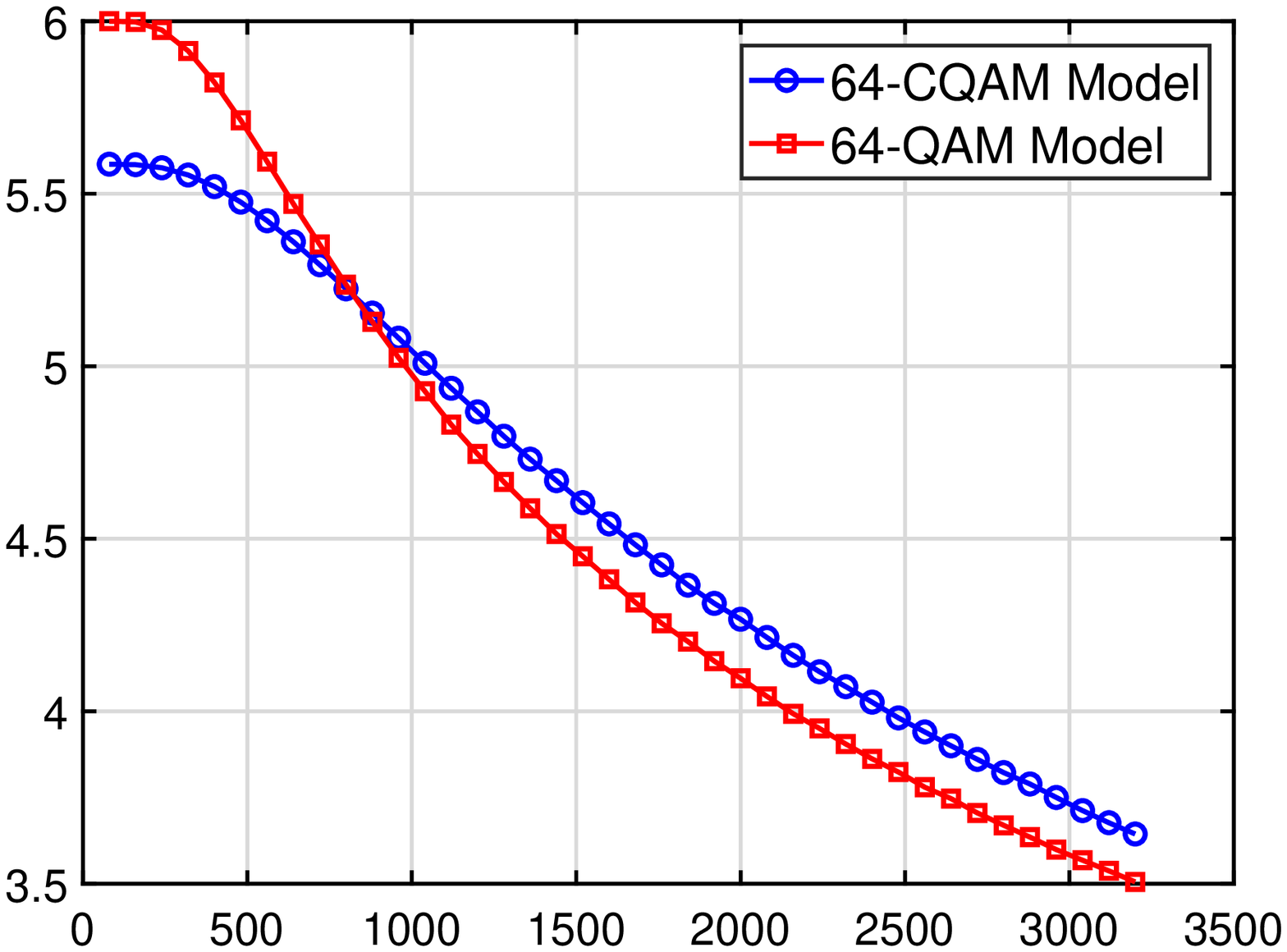} 
            \put(-240,30){\rotatebox{90}{\footnotesize MI at optimum SNR [bit/2D-symbol] }}
            \put(-140,-12){\rotatebox{0}{\footnotesize Transmission Distance [km]}}}      
   \caption{\footnotesize Modeling transmission distance results.
   \label{NL}}
       
        \end{minipage} 

\end{figure*}

\begin{figure*}[!t]

    \begin{minipage}[h]{0.5\linewidth}  
    \hspace{-0.15cm}
         \put(20,5){\includegraphics[width=0.9\linewidth]{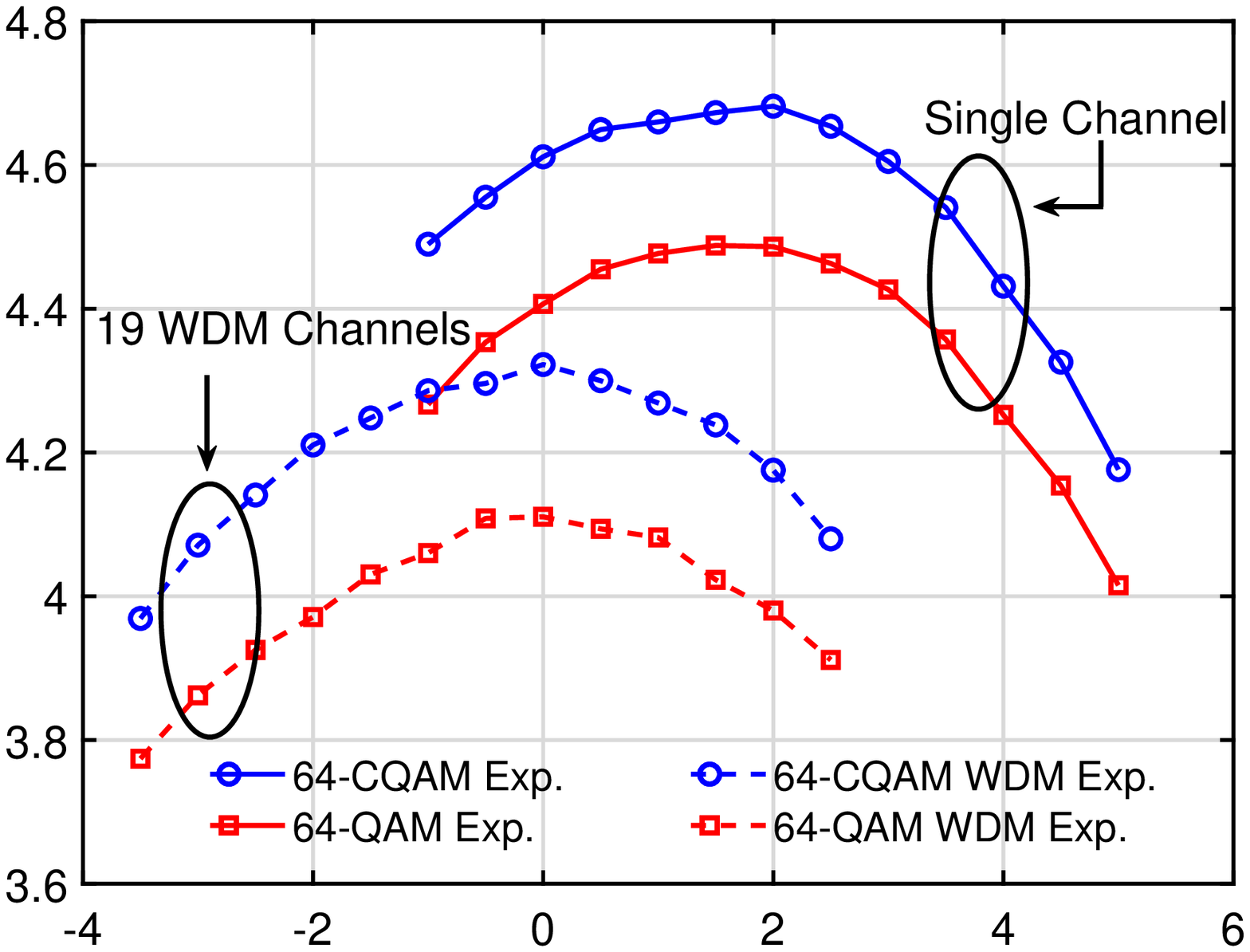}
          \put(-240,60){\rotatebox{90}{\footnotesize  MI [bit/2D-symbol]}}
         \put(-140,-10){\rotatebox{0}{\footnotesize Launch Power [dBm] }}}
        \subcaption{\footnotesize Exp. launch power results. \hspace{-1cm}\label{power}}
        \end{minipage} \hfill
         \begin{minipage}[h]{0.5\linewidth}  
         \hspace{-0.15cm}
        \put(10,5){\includegraphics[width=250pt]{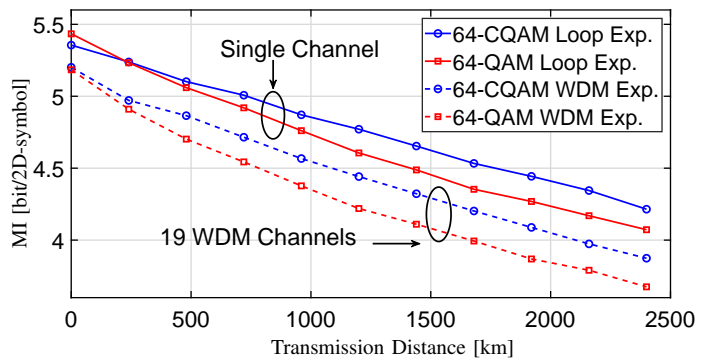}
        \put(-258,30){\rotatebox{90}{\footnotesize  MI [bit/2D-symbol]}}
         \put(-160,-10){\rotatebox{0}{\footnotesize Transmission Distance [km] }}}
        \subcaption{\footnotesize Exp. transmission distance results. \hspace{-1cm}  \label{distance}}
        \end{minipage}

\caption{\footnotesize 
(a) CM information rate as a function the launch power at 1440 km. (b) CM  information rate as a function of the transmission distance at the optimal launch power for single-channel and WDM transmissions.}
   
\end{figure*}
\section{Results} \label{sec:results}

The back-to-back results for 54.2 Gbaud are shown in Fig.~\ref{b2b} together with theoretical results. 
At a target mutual information (CM) of 4.5 bits/symb., we measure a 1.25 dB gain for 64-CQAM over 64-QAM. The target CM has been chosen for illustration purpose but still lies in the same region as the target CM of 4 bits/symb. of the previous sections. 
We note that 64-CQAM has a 0.7 dB lower implementation penalty compared to 64-QAM. This is most likely due a more efficient use of the DAC resolution when shaping is applied, see, e.g.,\cite{Buchali2017}. Notice that the experimental CM values have been determined knowing the transmitted sequence of the channel under test. This enables to build the signal statistic (estimated input distribution) and the channel model (estimated conditional distribution) associated with the experimental results up to some negligible (quantization) errors while the nonlinear Kerr-effects are treated as white Gaussian noise.

Fig.~\ref{power} shows the information rate $I(X;Y)$ (CM) as a function of the launch power at 1440 km for the two formats. We observe no apparent difference in the optimal launch power for the two formats in neither single channel nor WDM transmission. The optimal launch power per channel was around 2~dBm for single channel and 0~dBm for WDM.
The transmission results are depicted in Fig.~\ref{distance} for the optimal launch power. 
Assuming CM at 4.5 bits/symb., 54.2 Gbaud 64-CQAM can be transmitted up to 1750 km in single-channel transmission at the optimal launch power, and 1100 km with 19 WDM channels transmission.
Considering 19 WDM channels, if the formats are compared at CM = 4 bits/symb., the transmission distance can
be increased by 480 km by using 64-CQAM which corresponds to an increase of 28\%.

In the experiments, 64-CQAM has a slightly lower implementation penalty compared to 64QAM. For the shaped format, {\em clipping} is performed at the DAC level. Both formats suffer equally from hardware restrictions due in particular to the non-optimized evaluation board.
In order to verify the gains we see in experiments, without being influenced by the implementation penalties, we computed the mutual information of both formats, using formulas for the total variance of nonlinear distortions, which includes the impact of modulation format in the nonlinear regime, see Eq.~(123) in~\cite{Ghazisaeidi2017}. The transmitter and receiver are assumed ideal without implementation penalty, and the receiver DSP consists only of the matched filter. The modeling transmission results are depicted Fig.~\ref{NL}. At each distance, first the maximum SNR at optimum launch power is computed, then the corresponding optimum mutual information is computed. Fig.~\ref{NL} illustrates the optimum mutual information vs distance for 64-CQAM and 64-QAM. 64-CQAM has a clear advantage over 64-QAM beyond 1500 km. 
At CM = 4 bits/symb., the transmission reach can be increased by 14\% by switching from 64-QAM to 64-CQAM.


\section{Conclusion}

Interest in circular QAM emanates from a better matching to the polarization-multiplexed WDM fiber model, from the adaptation to non-binary processing, or from other evolved design constraints such as, potentially,  flexible rate adaptation for PAS.

Long-haul transmission simulations for shaped CQAM have indeed been compared to simulations for shaped 64-QAM in both the linear and nonlinear regime. Importantly, advanced simulations show that the new schemes have similar performance to the state-of-the-art schemes based on shaped QAM. Transmission experiments and comparisons with standard (unshaped) 64-QAM have validated this design and the use of CQAM for practical purpose. 
For example, in WDM transmission of 54.2~Gbaud signals, 64-CQAM achieved 28\% gain in transmission reach over conventional 64-QAM. 

 This work demonstrates that advanced shaping schemes such as combined geometric-probabilistic CQAM could be used and may have very interesting performance in practice. Assuming that significant performance gains result from advanced channel modeling and particular constellation geometry, and assuming that coding and modulation can be efficiently translated in high-speed transceivers, this may turn out to be key for the next generation of optical systems.

\section*{Acknoledgments}
The authors would like to thank L. Schmalen, A. Dumenil, and R.J. Essiambre for valuable comments and suggestions on an early version of this work. The authors are also grateful to the anonymous reviewers for their insightful and valuable comments.

\appendix

\subsection{Achievable Information Rates} \label{app:rate}

\begin{figure*}[h]
\begin{center}
\includegraphics[width=1.4\columnwidth]{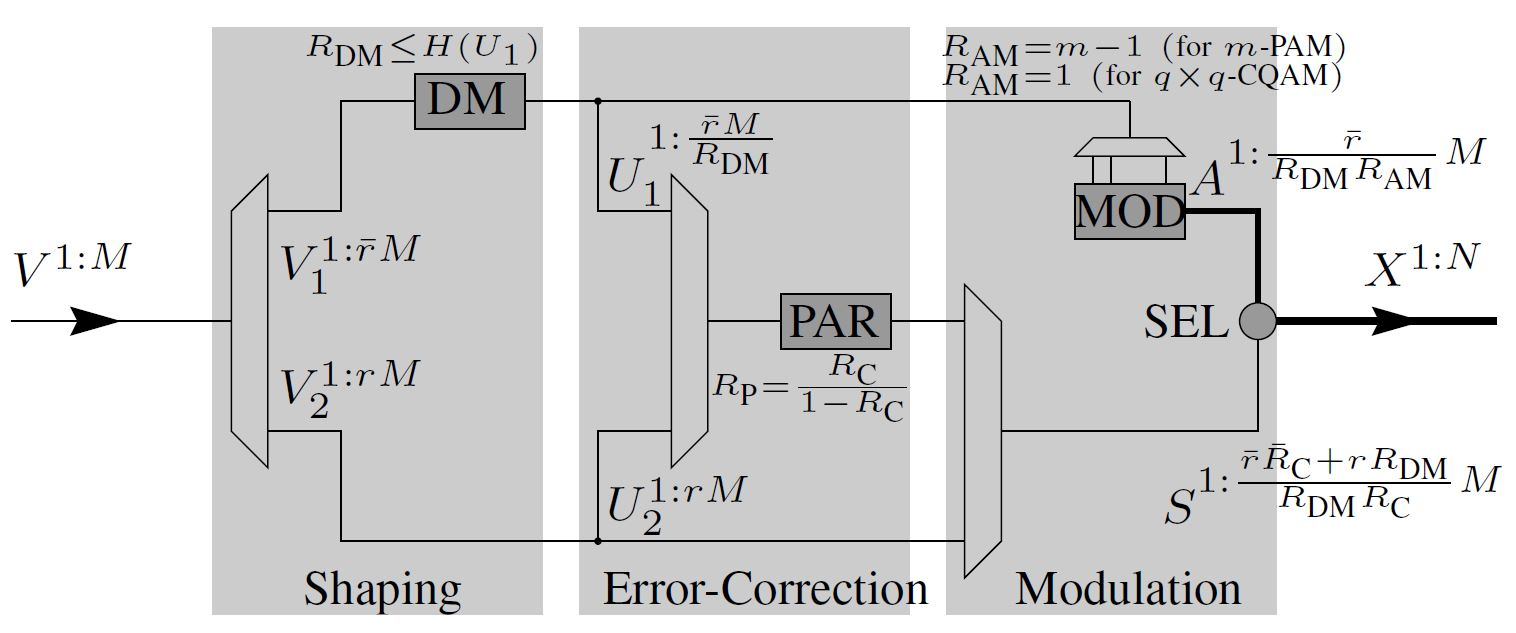}
\caption{\footnotesize Concatenation chain for PAS encoding. All operations up to final modulation are performed in the original symbol alphabet (thin lines) of size $|{\cal{V}}|=q$. The  task of the left-most encoding layer is {\em shaping}. To this aim, a standard source code (called {\em Distribution Matcher}) with rate $R_{\text{DM}}\in (0,1)$  is used after an initial split of the information sequence at rate $r\in[0,1)$ to further let the next layer be constraint-free. A peculiarity of PAS is to perform the shaping task up front conventional {\em error-correction} encoding. Error-correction encoding with rate $R_{\text{C}}\in(0,1)$ is then performed at the middle layer and generate a redundancy sequence (e.g., from linear parities). The right-most encoding layer is the {\em modulation} step: the general principle of PAS  is to select a region according to the (typically uniform) distribution  of the $S$s and combine it with a fundamental point labeled at rate $R_{\text{AM}}\geq 1$ with the (previously shaped) information sequence. Notice $U_2=V_2$.}
\label{fig:GPAS}
\end{center}
\end{figure*}

For a memoryless channel model with random input letters $X$ taking on discrete values $x\in\A$ with probability $p_X(x)$ at each channel use, the channel capacity is given by the information rate $I(X;Y)$. For the sake of simplicity, the term of CM (Coded Modulation) capacity is employed in this paper to refer to $I(X;Y)$ when the input alphabet $\A$ is fixed. For the complex-valued AWGN model, it is as a function of the SNR (ratio between the average constellation power and the additive noise). 
Modern error-correcting codes closely approach the achievable bounds in practical setups and for large blocklengths. See \cite{Merhav1994, Ganti2000} for capturing potential additional transceiver mismatches as well as \cite{Zehavi1992,Caire1998,Wachsmann1999,Guillen2008,Martinez2009} for operational characterizations. For the modulation schemes of our running examples involving square or circular QAM constellations, each letter modulates $m$ symbols of the original information alphabet represented by $S_1, S_2,\cdots,S_m$. In other words, there is a one-to-one mapping $\mu$ such that $X=\mu(S_1,\cdots,S_m)$. Notice that $m=3$ for $8$-PAM using binary labels or $m=2$ for $8\times8$-QAM constellations using $q=2^3$-ary labels. Using the chain rule and because conditioning reduces entropy, we see that 
 \begin{align*}
I(X;Y) & = H(X)-H(X|Y) \\ & \geq  H(S_1,\cdots,S_m)-\sum_{i=1}^m H(S_i|Y),
\end{align*}
i.e., the CM capacity is never less than the rate $(H(S_1,\cdots,S_m)-\sum_{i=1}^m H(S_i|Y))^+$ that indicates the system capacity when a {\em maximum a posteriori} (MAP) estimator operating at the {\em symbol level} is implemented (S-CM capacity). Notice that this expression encompasses the general case of correlated $S_i$s. By iterating the decomposition with $S_i=S_i(B_1^i,\cdots,B_\ell^i)$ for example, we also see that the capacity associated with symbol-MAP decoding (S-CM capacity) is not less than the capacity associated with bit-MAP decoding (B-CM capacity) provided that a symbol is labeled by a group of bits. Let us make a couple of observations. First, in the specific example $m=2$ of two symbols, the difference between the two is given by $I(S_1;S_2|Y)$ which, in the CQAM case or in \cite{Essiambre2010}, 
differentiates between amplitude and phase. Second, for independent symbols, the  capacity associated with symbol-MAP decoding (sometimes called bit-metric decoding \cite{Guillen2008}) can be written as $\sum_{i=1}^m I(S_i;Y)$. Hence, in this case, we may as well use the framework of ICM (Interleaved Coded Modulation) to define the notion of  achievable rates for conventional processing. The conceptual view of an infinite interleaver \cite{Zehavi1992,Caire1998} before any alphabet mapping and in conjunction with uniform signaling indeed permits to characterize different system capacities. More generally, it may be convenient to use the GMI framework in \cite{Merhav1994, Ganti2000} where the achievable rates permit to characterize conventional processing mismatches and have therefore an operational meaning. For the sake of clarity, we explicitly characterize the rate as CM or S-CM depending upon the choice of system architecture among those considered in this paper.

\subsection{Probabilistic Amplitude Shaping} \label{app:pas}

PAS originally stands for {\em Probabilistic Amplitude Shaping} \cite{Kramer2016}, a method devised in \cite{Calderbank1990, Bocherer2015} to implement non-uniform signaling \cite{Kschischang1993}. Although, in the presented generalized version, PAS no longer refers to modulating ``amplitudes'' as such, the original name is conserved for simplicity.

Assume that we want to communicate messages through $N$ independent uses of a communication channel. More precisely, let us denote by $({\cal{X}},p_X)$ the channel input where a random variable $X$ takes on values $x\in{\cal{X}}$ according to $p_X(x)$. For a source of independent symbols $V\in{{\cal{V}}}$ distributed uniformly at random ($|{{\cal{V}}}|=q$), the number of messages scales as ${\cal{M}}_M\defas|{\cal{V}}|^M=q^M$ where $M$ is the information length.  PAS \cite{Bocherer2015} is a layered coding scheme that maps the information symbols into the $X$s. The overall coding rate\footnote{When not stated otherwise, coding rate, information rate, or entropy are defined using $\log_q$ where $q$ is the original field characteristic.} is $R_T\defas\frac{\log_q({\cal{M}}_M   )}{N}=\frac{M}{N}$ where $N$ is the encoder output length.


The basic principle (amplitude sign flipping triggered by a bit when $q=2$) of the PAS method as devised in \cite{Bocherer2015} relies on binary channel symmetry. It is then tailored to binary-input real-valued-output symmetric channels such as the $2^m$-PAM AWGN channel (or product of it such as $2^{2m}$-QAM AWGN). It can be extended to $q$-ary channel symmetry and the generalized scheme is summarized\footnote{The notation $\bar{\rho}$ for $\rho\in[0,1]$ indicates the complement to one,  $\bar{\rho}\defas1-\rho$.} in Fig.~\ref{fig:GPAS}. 

\subsubsection{Sufficient Constellation} PAS relies on the subdivision of the input alphabet ${\cal{X}}$ into $J$ constellation regions $\{{\cal{C}}_j\}_{j\in\{1,\cdots,J\}}$ such that ${\cal{X}}=\cup_{j\in\{1,\cdots,J\}} {\cal{C}}_j$. Various constellations and subdivisions are PAS-compatible. For simplicity, assume that all constellation regions have same cardinality, i.e., $|{\cal{C}}_j|=|{\cal{C}}_1|$, with $\prob\{{\cal{C}}_j\} = \frac{1}{J}$ for any $j$. Assume further that $q=|{\cal{V}}|$ divides  
  $J$ and that $q=|{\cal{U}}_1|$ divides the region cardinality $|{\cal{S}}_1|$. For each region, the points are identically distributed. PAS is eventually performed on a reduced fundamental region $B$ chosen for example to be $B={\cal{C}}_1$. PAS coding consists in mapping labels obtained from  the sequence of (uniformly distributed) symbols of type $S$ into regions. Independently, PAS coding maps labels obtained from the (shaped) information sequence of type $U_1$ into fundamental points. In other words, there is a $m$ such that $\A\equiv \{0,1,\cdots,q-1\}^m\times B$, with $B\defas {\cal{C}}_1$.    


\subsubsection{State-of-the-Art and Legacy Systems} 
Let us provide here some background on coding in actual optical applications. In practice, forward error-correction (FEC) is typically performed via a (systematic) linear code of rate $R_\text{C}$. PAS is said to be compatible with legacy systems because it can be built around a standard (or pre-existing) FEC coding engine (e.g., an LDPC-based system). 
 PAS first focuses on shaping the distribution of points inside the fundamental region using the {\em distribution matcher} (DM). To exemplify this, let us use the binary case $q=2$. The distribution of the $2^m$-PAM amplitudes is shaped to let the distribution of the full constellation behave like the capacity-achieving Gaussian \cite{Kschischang1993}. If the standard PAM modulation rate is $R_{\text M}=m$, then PAS modulates the signal amplitudes at the output of the distribution matcher at rate $R_{\text{AM}}=m-1$. PAS uses (up to very few operational changes) a conventional coding and modulation chain. After the DM, the information sequence is parsed to modulate the point amplitudes while the (uniformly distributed) parity bits (as well as the unshaped information fraction) encode the sign of the PAM amplitudes.  The binary case is used for example in \cite{Ghazisaeidi2016}.

\subsubsection{General Framework}
As depicted in  Fig.~\ref{fig:GPAS}, PAS is seen as a layered coding system. The concatenation chain is divided into three main layers and encoding operations are done in a sequential order. Practical decoding is envisioned to occur in the reversed order. A fraction $\bar{r}\defas 1-r$ (with in some cases $r=0$) of the information stream is first encoded into a sequence (seen as a sequence of symbol packets or labels) with a given (required) distribution (typically Maxwell-Boltzmann as in \cite{Kschischang1993}). Hence, independent identically uniformly distributed symbols are encoded into a symbol sequence which (from parsing) labels the modulated regions at rate $R_{\text{AM}}$. The rate $R_{\text{AM}}$ is equal to the number of symbols in an alphabet of size $q$ needed to label a region (for example, $R_{\text{AM}}=m-1$ in the binary case of \cite{Bocherer2015} where a region is an amplitude, or $R_{\text{AM}}=1$ in the non-binary case of \cite{Boutros2017}). Second, a sequence of redundant symbols, generally obtained from linear combinations of information symbols, is then generated by a linear channel encoder. Dense linear combinations of symbols make that the distribution of resulting sum symbols tends to get asymptotically uniform. Third, the final encoding layer modulates symbols in ${\cal{X}}$ by selecting a pair composed of a point (for example representing an amplitude) in the fundamental region according to the label sequence (for example representing a quadrant or an angular region). 

\subsubsection{Compatible Rates}
A compatibility criteria of the set of rates $\{r,R_{\text{DM}},R_{\text{C}}, R_{\text{AM}}\}$ is easily obtained from Fig.~\ref{fig:GPAS}. Consider the layered encoding flow. We see that the system is constrained at the selective 
node when the end (modulation) layer is processed. The constraint reads $N= \frac{\bar{r}}{R_{\text{DM}}R_{\text{AM}}}M=\frac{\bar{r}\bar{R}_{\text{C}} +r R_{\text{DM}}}{R_{\text{DM}}R_{\text{C}}}M$. Its satisfaction implies a dependency between the rates as $R_{\text{C}}-r R_{\text{C}}-R_{\text{AM}}+R_{\text{AM}}R_{\text{C}}+r R_{\text{AM}}-r R_{\text{AM}} R_{\text{C}} -r R_{\text{AM}} R_{\text{DM}}=0$. When solved for $R_{\text{C}}$, it shows that $R_{\text{C}}=\frac{R_{\text{AM}}}{1+R_{\text{AM}}}(1+\frac{r}{1-r}R_{\text{DM}})$, i.e., the choice of the core channel code may be restricted to particular code rates. A first example is the binary case with $2^m$-PAM for which it is required to have $R_{\text{C}}= \frac{m-1}{m}(1+\frac{r}{\bar{r}}R_{\text{DM}})\geq \frac{m-1}{m}$ (achieved for $r=0$). This translates as $R_{\text{C}}\geq \frac{1}{2}$ for $16$-QAM or $R_{\text{C}}\geq \frac{2}{3}$ for $64$-QAM (bit-triggered region selection \cite{Bocherer2015}). A second 
 example is the $q$-ary case with  $q^2$-CQAM for which $R_{\text{C}}= \frac{1}{2}(1+\frac{r}{\bar{r}}R_{\text{DM}})\geq \frac{1}{2}$ for any $q^2$-CQAM (symbol-triggered region selection \cite{Boutros2017}).

\subsubsection{PAS Information Rate}
The splitting rate $r$ 
provides the designer with the degree of freedom that is necessary to satisfy the rate constraint. When solved for $r$, the compatibility constraint gives $r=1-\frac{R_{\text{AM}} R_{\text{DM}}}{R_{\text{C}}-R_{\text{AM}}+R_{\text{AM}}R_{\text{C}}+ R_{\text{AM}} R_{\text{DM}}}$. Therefore the overall PAS coding rate is 
\begin{align*}
R_{\text{T}}  & = R_{\text{C}}-R_{\text{AM}}+R_{\text{AM}}R_{\text{C}}+ R_{\text{AM}} R_{\text{DM}}.
\end{align*}
In our binary and non-binary running examples, this gives 
 $R_{\text{T}}  = m R_{\text{C}} - (m-1)(1- R_{\text{DM}})$ for $m$-PAM-based schemes and $R_{\text{T}}  = 2 R_{\text{C}} - 1 + R_{\text{DM}}$ for $q^2$-CQAM-based schemes, respectively. Expressed in binary units, those rates express the operational spectral efficiency of the respective coding systems. For example, for the constellations of two real dimensions and $2^m$ points of our running examples, the respective system capacities in bits per channel uses are  
\begin{align*}
R^b_{\text{T}}  & = 2 m R_{\text{C}} - 2 (m-1)(1- R_{\text{DM}})
\end{align*}
 for $2^{m}\times 2^m$-QAM (bit-triggered) and 
$R_{\text{T}}^b  = 2 \log_2(q) R_{\text{C}} - \log_2(q)(1 - R_{\text{DM}})$
 for $q^2$-CQAM (phase-triggered), i.e., 
\begin{align*}
R_{\text{T}}^b  =  2 m R_{\text{C}} - m (1- R_{\text{DM}})
\end{align*}
for $(2^{m})^2$-CQAM. Notice that the maximal transmitted entropy is $H(X)=H(V_1)+H(S)$ as the region and points within the fundamental region are independent. For our running examples, we see that the binary entropy becomes $H(X)\leq \log_2(q) R_{\text{DM}} + \log_2(q)\leq 2 \log_2(q)$. This represents the maximal amount of information that  PAS may transmit.

\newcommand{\ieeeit}{{\em IEEE Trans.~Inf.~Theory}}
\newcommand{\ieeecom}{{\em IEEE Trans.~Commun.}}
\newcommand{\ieeesac}{{\em IEEE J.~Sel.~Areas~Commun.}}


\begin{thebibliography}{1}


\bibitem{Shannon1948}  C.E.~Shannon, 
``A Mathematical Theory of Communications,'' 
{\em The Bell Technical System Journal,} 
Vol. 27, Issue 3, July 1948. 

\bibitem{Gallager1968} 
R.~G.~Gallager, 
{\em Information theory and reliable communication}. 
New York: Wiley, 1968.

\bibitem{Ungerboeck1982}
 G. Ungerboeck, 
``Channel coding with multilevel/phase signals,'' 
\ieeeit, 
vol.~28, pp.~55--67, Jan.~1982.

\bibitem{Calderbank1987} 
A.~R.~Calderbank and N.~J.~A.~Sloane, 
``New trellis codes based on lattices and cosets,''
\ieeeit,
vol.~33, no.~2, pp.~177--195, Mar.~1987.

\bibitem{Forney1989a} 
G.~D.~Forney and L.-F.~Wei, 
``Multidimensional constellations -- Part I: Introduction, figures of merit, and generalized cross constellations,''
\ieeesac, vol.~7, no.~6, pp.~877--892, Aug.~1989.


\bibitem{Calderbank1990} 
A.~R.~Calderbank and L.~H.~Ozarow, 
``Non-equiprobable signaling on the Gaussian channel,''
\ieeeit, 
vol.~36, no.~4, pp.~726--740, Jul.~1990.

\bibitem{Fortier1992} 
P.~Fortier, A.~Ruiz, and J.~M.~Cioffi, 
``Multidimensional signal sets through the shell construction for parallel channels,'' 
\ieeecom,
vol.~40, no.~3, pp.~500--512, Mar.~1992.

\bibitem{Forney1992} 
G.~D.~Forney, ``Trellis shaping,'' 
\ieeeit, 
vol.~38, no.~2, pp.~281--300, Mar.~1992.

\bibitem{Khandani1993} 
A.~K.~Khandani and P.~Kabal, 
``Shaping multidimensional signal spaces -- Part 1. Optimum shaping, shell mapping,''
\ieeeit,
vol.~39, no.~6, pp.~1799--1808, Nov.~1993.

\bibitem{Laroia1994} 
R.~Laroia, N.~Farvardin, and S.~A.~Tretter, 
``On optimal shaping of multi-dimensional constellations,''
\ieeeit,
vol.~40, no.~4, pp.~1044--1056, Jul.~1994.

\bibitem{Betts1994} W. Betts, A. R. Calderbank, and R. Laroia, 
``Performance of Nonuniform Constellations on the Gaussian Channel,'' 
\ieeeit, vol.~40, pp. 1633--1638, Sep. 1994.

\bibitem{Kschischang1993} 
F.~R.~Kschischang and S.~Pasupathy, 
``Optimal Nonuniform Signaling for Gaussian Channels,'' 
\ieeeit, 
vol.~39, no.~3, pp.~913--929, May 1993.

\bibitem{Forney2000} 
G.~D.~Forney, M.~D.~Trott, and S.-Y.~Chung,
``Sphere-bound-achieving coset codes and multilevel coset codes,'' 
\ieeeit, 
vol.~46, no.~3, pp.~820--850, May~2000.

\bibitem{Erez2005} 
U.~Erez, S.~Litsyn, and R.~Zamir,
``Lattices which are good for (almost) everything,'' 
\ieeeit,
vol.~51, no.~10, pp.~3401--3416, Oct.~2005.


\bibitem{Buda1989} R.~de~Buda. 
``Some optimal codes have structure.'' 
\ieeesac,  
vol.~7, no.~6, pp.~893--899, Aug.~1989.

\bibitem{Boutros1996} J. Boutros, E. Viterbo, C. Rastello, and J.-C. Belfiore, 
``Good lattice constellations for both Rayleigh fading and Gaussian channels,'' 
\ieeeit, 
vol.~42, no.~2, pp.~502--518, Mar.~1996.

\bibitem{Loeliger1997} H.A.~Loeliger, 
``Averaging bounds for lattices and linear codes,'' 
\ieeeit, 
vol~.43, no.~6, pp.~1767--1773, Nov.~1997.

\bibitem{Berrou1993}
 C. Berrou, A. Glavieux, and P. Thitimajshima, 
``Near Shannon limit error correcting coding and decoding,' 
{\em ICC,} Geneve, Switzerland, pp.~1064--1070, May 1993.

\bibitem{Gallager1963}
 R.~G.~Gallager, 
{\em Low-Density Parity-Check codes.} 
Cambridge, MA: MIT Press, 1963.


\bibitem{Caire1998}
G.~Caire, G.~Taricco, and E.~Biglieri,
``Bit-Interleaved Coded Modulation,''
\ieeeit,
vol.~44, no.~23, pp.~927--946, May~1998.

\bibitem{Imai1977} H. Imai and S.Hirakawa, 
``A multilevel coding method using error-correcting codes,''
\ieeeit, vol.~23, pp.~371--377, 1977.

\bibitem{Zehavi1992} E. Zehavi, ``8-PSK trellis codes for a Rayleigh channel,'' 
\ieeecom, 
vol.~40, no.~5, pp.~873--884, May~1992.


\bibitem{McEliece2001} R. J. McEliece, ``Are turbo-like codes effective on nonstandard channels?''
 {\em IEEE Inform. Theory Soc. Newslett.,} vol.~51, no.~4, pp.~1--8, Dec. 2001.

\bibitem{Soriaga2003} J. B. Soriaga and P. H. Siegel, 
``On distribution shaping codes for partial-response channels,''
 {\em Allerton Conf. on Commun., Control, and Computing, Monticello,}  USA, Oct. 2003.

\bibitem{Gabrys2012} R. Gabrys and L. Dolecek, ``Coding for the Binary Asymmetric Channel,'' {\em Int. Conf. on Computing Networking and Communications,} pp.~461--465, 2012.

\bibitem{Ling2014} C. Ling and J.C. Belfiore, 
``Achieving AWGN channel capacity with lattice Gaussian coding,'' 
\ieeeit, vol.~60, no.~10, pp.~5918--5929, Oct. 2014.

\bibitem{Mondelli2014} M. Mondelli, S.H. Hassani, and R. Urbanke, 
``How to Achieve the Capacity of Asymmetric Channels,'' 
{\em Allerton Conf. on Commun., Control, and Computing, Monticello,} pp.~789--796, Oct. 2014.


\bibitem{Palgy2012} 
N. Palgy and R. Zamir. 
``Dithered probabilistic shaping,'' 
{\em In IEEE 27th Convention of Electrical Electronics Engineers in Israel,} Nov. 2012.

\bibitem{Bocherer2015}
G.~B{\"o}cherer, F.~Steiner,and P.~Schulte,
``Bandwidth Efficient and Rate-Matched Low-Density Parity-Check Coded Modulation,''
\ieeecom,  
vol.~63, no.~12, pp.~4651--4665, Dec.~2015.

\bibitem{Schulte2016} 
P.~Schulte and G.~B\"ocherer, 
``Constant Composition Distribution Matching,'' 
\ieeeit, 
vol.~62, no.~1, pp.~430--434, Jan.~2016.

\bibitem{Kramer2016} 
G.~Kramer, 
``Probabilistic amplitude shaping applied to fiber-optic communication systems,'' 
{\em Int.~Symp.~on~Turbo~Codes~and~Iterative~Inf.~Proc.,} Oct. 2016.

\bibitem{Boutros2017}
J.J.~Boutros, F.~Jardel, and C.~Measson,
``Probabilistic Shaping and Non-Binary Codes,''
\emph{ ISIT}, pp.~2308-2312, Jun. 2017.

\bibitem{Yankov2014} M.P. Yankov, D. Zibar, K.J. Larsen, L.P.B. Christensen, and S. Forchhammer, ``Constellation Shaping for Fiber-Optic Channels with QAM and High Spectral Efficiency,'' 
{\em IEEE Photon.~Technol.~Lett.}, 
vol.~26, no.~23, pp.~2407--2410, Dec. 2014.

\bibitem{Beygi2014}
L. Beygi, E. Agrell, J.M. Kahn, and M. Karlsson, ``Rate-Adaptive Coded Modulation for Fiber-Optic Communications,'' {\em IEEE J. Lightwave Technol.},vol.~32, no.~2, pp.~333--343, Jan.~2014.

\bibitem{Fehenberger2015}
T. Fehenberger, G. B\"ocherer, A. Alvarado, and N. Hanik, ``LDPC coded modulation with probabilistic shaping for optical fiber systems,'' \emph{OFC}, paper Th2A.23, Mar. 2015.

\bibitem{Buchali2015} F. Buchali, G. B{\"o}cherer, W. Idler, L. Schmalen, P. Schulte, F. Steiner, ``Experimental Demonstration of Capacity Increase and Rate-Adaptation by Probabilistically Shaped 64-QAM,'' \emph{ECOC}, Sep. 2015.


\bibitem{Ghazisaeidi2016} A. Ghazisaeidi, I. Fernandez de Jauregui, R. Rios-Mueller, L. Schmalen, P. Tran, P. Brindel, A. Carbo Meseguer, Q. Hu, F. Buchali, G. Charlet, and J. Renaudier, ``65Tb/s Transoceanic Transmission using Probabilistic Shaping,'' \emph{ECOC}, Sep. 2016.

\bibitem{Chandrasekhar2016} S. Chandrasekhar, B. Li, J. Cho, X. Chen, E. Burrows, G. Raybon, P. Winzer, ``High-spectral-efficiency transmission of PDM 256-QAM with Parallel Probabilistic Shaping at Record Rate-Reach Trade-offs,'' \emph{ECOC}, Sep. 2016.

\bibitem{Zhang2016} S. Zhang, F. Yaman, Y.K. Huang, J.D. Downie, D. Zou, W. A. Wood, A. Zakharian, R. Khrapko, S. Mishra, V. Nazarov, J. Hurley, I.B. Djordjevic, E. Mateo, and Y. Inada,  ``Capacity-Approaching  Transmission over  6375  km  at  Spectral  Efficiency  of  8.3  bit/s/Hz,''  \emph{OFC}, paper Th5C.2, Mar. 2016.

\bibitem{Theresa2017}
Q. Hu, F. Buchali, L. Schmalen, and H. Buelow, ``Experimental Demonstration of Probabilistically Shaped QAM,''  \emph{Advanced Photonics 2017, OSA Technical Digest (online)}, paper SpM2F.6, 2017. 

\bibitem{Ivan2017}
I. F. de Jauregui Ruiz, A. Ghazisaeidi, R. Rios-Muller, and P. Tran, ``Performance Comparison of Advanced Modulation Formats for Transoceanic Coherent Systems,'' \emph{OFC}, paper Th4D.6, 2017.

\bibitem{Jardel2017} F. Jardel, T. Eriksson, F. Buchali, W. Idler, A. Ghazisaeidi, C. M{\'e}asson, and J. Boutros, ``Experimental Comparison of 64-QAM and Combined Geometric-Probabilistic Shaped 64-QAM,'' {\em ECOC}, Tu.1.D.5, Sep. 2017.

\bibitem{Buchali2017}
F. Buchali, et al., ``Flexible Optical Transmission close to the Shannon Limit by Probabilistically Shaped QAM,'' \emph{in Proc. OFC}, paper M3C.3, Mar. 2017.

\bibitem{Agrawal2012} 
G.P.Agrawal, 
``Nonlinear Fiber Optics,'' 
5-th Edition, {\em Academic Press}, Oct. 2012.

\bibitem{Mecozzi2000}
 A.~Mecozzi, C.B.~Clausen, and M.~Shtaif, 
``System impact of intrachannel nonlinear effects in highly dispersed optical pulse transmission,''
{\em IEEE Photon. Tech. Lett.}, vol. 12, no. 12, pp. 1633--1635, Dec. 2000.

\bibitem{Dar2013}
R.~Dar, M.~Feder, A.~Mecozzi, and M.~Shtaif, 
``Properties of nonlinear noise in long dispersion-uncompensated fiber links,''
{\em Optics Express}, 
vol. 21, no. 22, pp. 25685--25699, Oct. 2013.

\bibitem{Carena2014} 
A. Carena, G. Bosco, V. Curri, Y. Jiang, P. Poggiolini, and F. Forghieri,
``EGN model of nonlinear fiber propagation,'' 
{\em Optics. Express,} vol. 22, no. 13 pp. 16335--16362, May 2014.

\bibitem{Eriksson2016}
T. Eriksson, T. Fehenberger, P. Andrekson, M. Karlsson, N. Hanik, and E. Agrell, ``Impact of 4D Channel Distribution on the Achievable Rates in Coherent Optical Communication Experiments,''  {\em IEEE J. Lightwave Technol.}, vol. 34, pp. 2256--2266 , May 2016.



\bibitem{Dar2016} R.~Dar, M.~Feder, A.~Mecozzi, and M.~Shtaif, ``Pulse Collision Picture of
Inter-Channel Nonlinear Interference in Fiber-Optic Communications,''
{\em IEEE J. Lightwave Technol.,} vol. 34, no. 2, pp. 593-607, Jan.~2016.

\bibitem{Ghazisaeidi2017}
A. Ghazisaeidi, ``A Theory of Nonlinear Interactions between Signal and Amplified Spontaneous Emission Noise in Coherent Wavelength Division Multiplexed Systems, " {\em IEEE J. Lightwave Technol.}, vol. 44, no. 23, pp. 5150--5175, Dec. 2017.

\bibitem{Awwad2013}
E~Awwad, Y.~Jaou{\"e}n and G.~Rekaya-Ben~Othman
``Polarization-time coding for PDL mitigation in long-haul PolMux OFDM systems,''
{\em Optics Express, OSA,} vol. 21, no. 19, pp. 22773--22790, 2013.

\bibitem{Dumenil2017}
A.~Dumenil, E.~Awwad, C.~M{\'e}asson,
``Polarization Dependent Loss: Fundamental Limits and How to Approach Them,''
{\em Signal Processing in Photonic Commun. Conf.,}
New Orleans, Louisiana, USA, Jul. 2017.

\bibitem{Dumenil2018}
A.~Dumenil, E.~Awwad, C.~M{\'e}asson,
``Low-Complexity Polarization Coding for PDL-Resilience,''
Accepted for publication, Th1F.5., 4071998, ECOC, Sep. 2018.

\bibitem{Yankovn2017}
M.P. Yankovn, F. Da Ros, E.P. da Silva, S. Forchhammer, K.J. Larsen, L.K. Oxenlowe, M. Galili, and D. Zibar,
``Constellation Shaping for WDM Systems Using 256QAM/1024QAM With Probabilistic Optimization,'' 
{\em IEEE J. Lightwave Technol.,} vol. 34, no. 22, pp. 5146-5156, Nov.~2015.

\bibitem{Renner2017}
J. Renner, T. Fehenberger, M.P. Yankov, F. Da Ros, S. Forchhammer, G. Böcherer, and N. Hanik,
``Experimental Comparison of Probabilistic Shaping Methods for Unrepeated Fiber Transmission,''
{\em IEEE J. Lightwave Technol.,} vol. 35, no. 22, pp. 4871-4879, Nov.~2017.

\bibitem{Pan2016}
C. Pan and F. R. Kschischang,
``Probabilistic 16-QAM shaping in WDM systems,'' 
{\em IEEE J. Lightwave Technol.,} vol. 34, no. 18, pp. 4285–4292, Jul.~2016.

\bibitem{Fehenberger2016}
T. Fehenberger, A. Alvarado, G. Bocherer, and N. Hanik, 
``On probabilistic shaping of quadrature amplitude modulation for the nonlinear fiber channel,'' 
{\em IEEE J. Lightwave Technol.,} vol. 34, no. 21, pp. 5063–5073, Jul.~2016.




\bibitem{Wachsmann1999}
U.~Wachsmann, R.~Fischer, and J.B.~Huber, 
``Multilevel codes: Theoretical concepts and practical design rules,''
\ieeeit,
vol.~45, no.~5, pp.~1361--1391, Jul.~1999.

\bibitem{Guillen2008}
A. Guill{\'e}n i F{\`a}bregas, A. Martinez, and G. Caire,
``Bit-Interleaved Coded Modulation,'' 
{\em Foundations and Trends® in Communications and Information Theory}, 
vol. 5, no.~1–2, pp.~1--153, 2008.

\bibitem{Martinez2009}
A. Martinez, A. Guill{\'e}n i F{\`a}bregas, G.~Caire, and F. Willems, 
``Bit-Interleaved Coded Modulation Revisited: A Mismatched Decoding Perspective,''
\ieeeit,
vol.~55, no.~6, pp.~2756--2765, Jun.~2009.

\bibitem{Merhav1994}
N.~Merhav, G.~Kaplan, A.~Lapidoth, and S. Shamai (Shitz), 
``On information rates for mismatched decoders,''
\ieeeit,
vol.~40, no.~6, pp.~1953--1967, Nov.~1994.

\bibitem{Ganti2000}
A.~Ganti, A.~Lapidoth, and E.~Telatar, 
``Mismatched decoding revisited: General alphabets, channels with memory, and the wideband
limit,''
\ieeeit, 
vol. 46, no. 7, pp. 2315--2328, Nov.~2000.


\bibitem{Essiambre2010}
R.J.~Essiambre, G.~Kramer, P.J.~Winzer, G.J.~Foschini, B.~Goebel,
``Capacity limits of optical fiber networks,''
{\em IEEE J. Lightwave Technol.},
vol. 28, no. 4, pp. 662--701, 2010.

\bibitem{Larsson2017} 
P.~Larsson, 
``Golden Angle Modulation,'' 
 \emph{submitted for pub. in IEEE Wireless Comm. Let.}, 
Sep. 2017.


%
%
%
%
%
%
%
%
%
%
%
%
%
%
%
%
%
%
%
%
%
%
%
\end{thebibliography}
\end{document}